\documentclass[aip,pop,superscriptaddress,showpacs]{revtex4-1}
\usepackage{amssymb,amsmath,amsfonts,graphicx,epsf,color}

\begin{document}

\title{Scalings of intermittent structures in magnetohydrodynamic turbulence}

\author{Vladimir Zhdankin\footnote{Invited speaker}}
\email{zhdankin@jila.colorado.edu}
\affiliation{JILA, NIST and University of Colorado, 440 UCB, Boulder, Colorado 80309, USA}
\author{Stanislav Boldyrev}
\affiliation{Department of Physics, University of Wisconsin-Madison, 1150 University Avenue, Madison, Wisconsin 53706, USA}
\affiliation{Space Science Institute, Boulder, Colorado 80301, USA}
\author{Dmitri A. Uzdensky}
\affiliation{Center for Integrated Plasma Studies, Physics Department, UCB-390, University of Colorado, Boulder, CO 80309} 

\date{\today}

\begin{abstract}
Turbulence is ubiquitous in plasmas, leading to rich dynamics characterized by irregularity, irreversibility, energy fluctuations across many scales, and energy transfer across many scales. Another fundamental and generic feature of turbulence, although sometimes overlooked, is the inhomogeneous dissipation of energy in space and in time. This is a consequence of intermittency, the scale-dependent inhomogeneity of dynamics caused by fluctuations in the turbulent cascade. Intermittency causes turbulent plasmas to self-organize into coherent dissipative structures, which may govern heating, diffusion, particle acceleration, and radiation emissions. In this paper, we present recent progress on understanding intermittency in incompressible magnetohydrodynamic turbulence with a strong guide field. We focus on the statistical analysis of intermittent dissipative structures, which occupy a small fraction of the volume but arguably account for the majority of energy dissipation. We show that, in our numerical simulations, intermittent structures in the current density, vorticity, and Els\"{a}sser vorticities all have nearly identical statistical properties. We propose phenomenological explanations for the scalings based on general considerations of Els\"{a}sser vorticity structures. Finally, we examine the broader implications of intermittency for astrophysical systems.
\end{abstract}


\maketitle

\section{Introduction}

Turbulence is generally regarded to be the complex spatiotemporal behavior of a dynamical field characterized by irregular and irreversible dynamics, fluctuations across many scales, and energy exchange between many scales. Although a formidable problem, the relative simplicity of the dynamical equations (e.g., the Navier-Stokes equations) and the presence of symmetries implies that a statistical framework for describing turbulence may in principle be achieved. In his original theory for incompressible hydrodynamic turbulence, Kolmogorov \citep{kolmogorov_1941} proposed that the inertial-range dynamics can be described by scale invariance, with the mean energy dissipation rate $\langle \epsilon \rangle$ and length scale $l$ being the only relevant variables. Dimensional analysis then constrains the moments of increments in the velocity field, known as structure functions, to scale as $\langle (\delta v_l)^n \rangle \propto (\langle \epsilon \rangle l)^{n/3}$ for all orders $n$, where longitudinal velocity increments are defined as $\delta v_l(\boldsymbol{x}) = [\boldsymbol{v}(\boldsymbol{x}+\boldsymbol{l})-\boldsymbol{v}(\boldsymbol{x})] \cdot \boldsymbol{l}/l$ with $\boldsymbol{l}$ in an arbitrary direction. The $n=2$ structure function implies an energy spectrum $E(k) \propto k^{-5/3}$, while the $n=3$ structure function gives Kolmogorov's four-fifths law\citep{frisch_1995}.

Although the predicted low-order statistics (associated with $n \le 3$) agree remarkably well with experiments and numerical simulations, observations show that higher-order statistics (associated with $n > 3$) deviate strongly from Kolmogorov's predicted scaling. The loophole in Kolmogorov's argument comes from a generic phenomenon known as intermittency, which causes $\langle \epsilon \rangle$ to be insufficient for characterizing the dynamics due to strong local fluctuations in dissipation.

Intermittency is the inherent spatiotemporal inhomogeneity of turbulence due to random fluctuations in the energy cascade as it proceeds from large scales to small scales. To illustrate intermittency, one can consider the local energy dissipation rate measured across a subvolume (e.g., sphere) of size $l$, denoted by $\epsilon_l$. Due to the development of the cascade across an increasing number of scales, $\epsilon_l$ spans a wider range of values as one considers smaller scales $l$. Hence, the probability density function for $\epsilon_l$ is scale-dependent, with the moments $\langle \epsilon_l^n \rangle$ increasing with decreasing scale. The shortcoming of Kolmogorov's original theory was that $\langle \epsilon \rangle$ was used when instead the random variable $\epsilon_l$ should be used\citep{kolmogorov_1962, obukhov_1962}, making a complete theory nontrivial to construct. 

One important consequence of intermittency is the emergence of coherent structures and intense dissipative events. In turbulent plasmas, intermittency forms current sheets and vorticity sheets that may serve as sites for magnetic reconnection, plasma heating, and particle acceleration; they may also impede particle transport and affect the magnetic dynamo. Intermittency may play an important role in fusion devices \citep{carbone_etal_2000, antar_etal_2003, dippolito_etal_2004}, the solar photosphere, corona, and wind \cite{cattaneo_1999, bushby_houghton_2005, stein_nordlund_2006, cranmer_etal_2007, osman_etal_2011, uritsky_etal_2007}, Earth's magnetosphere \citep{angelopoulos_etal_1999, uritsky_etal_2002}, the interstellar medium \cite{kowal_etal_2007, pan_etal_2009}, accretion systems \cite{dimatteo_etal_1999, eckart_etal_2009, albert_etal_2007}, and pulsar wind nebulae \cite{tavani_etal_2011, abdo_etal_2011}. For a recent review on intermittency in plasmas, see Matthaeus et al. 2015\citep{matthaeus_etal_2015}.

A promising methodology for understanding intermittency is the statistical analysis of intermittent structures. The energetics, sizes, and morphology of intermittent structures can reveal inhomogeneity, characteristic dynamical scales, and anisotropy. The properties of the structures can also be used for modeling physical processes such as heating, transport, particle acceleration, and radiation emission. Intermittent structures are identifiable not only in numerical simulations, but also in a large class of experimental and observational problems, although usually with limitations such as reduced dimensionality (e.g., 1D temporal measurements from spacecraft in the solar wind or projected 2D spatial emission profiles from the solar corona) or lack of direct measurements of key quantities (e.g., the local energy dissipation rate, which must be inferred from proxies). Methods were successfully developed and applied for the quantitative statistical analysis of intermittent vorticity filaments in numerical simulations of hydrodynamic turbulence \citep{jimenez_etal_1993, jimenez_wray_1998, moisy_jimenez_2004, leung_etal_2012}. On the other hand, although intermittent structures were long known to exist in numerical simulations of MHD turbulence\citep[e.g.,][]{politano_etal_1995b, muller_biskamp_2000, muller_etal_2003, biskamp_2003}, methods to study them in substantial detail were developed only recently. Statistical analyses were performed to understand the role of coherent structures in the kinematic dynamo \citep{wilkin_etal_2007}; magnetic reconnection in 2D MHD turbulence \citep{servidio_etal_2009, servidio_etal_2010} and reduced MHD turbulence \citep{zhdankin_etal_2013, wan_etal_2014}; dissipation in decaying MHD turbulence \citep{uritsky_etal_2010, momferratos_etal_2014} and driven MHD turbulence \citep{zhdankin_etal_2014, zhdankin_etal_2015, zhdankin_etal_2015b}; and current sheets in decaying collisionless plasma turbulence \cite{makwana_etal_2015}.

In our recent work on dissipative structures in MHD turbulence, we found that a significant fraction of the resistive energy dissipation is concentrated in intermittent current sheets\citep{zhdankin_etal_2014}. These current sheets are large in the sense that, although very thin, their widths and lengths span inertial-range scales up to the system size. In fact, the contribution of the largest current sheets to the overall energy dissipation is comparable to, if not greater than, that from dissipation-scale current sheets. In this work, we expand upon those conclusions and compare the statistical properties of the current sheets to dissipative structures in the vorticity and Els\"{a}sser vorticity fields. We show that the statistics are nearly identical in each of the three intermittent fields. We also discuss how the reduced MHD equations for the Els\"{a}sser vorticities yield insight into the observed scalings of intermittent structures. Finally, we conclude with an overview of some astrophysical implications of intermittent structures and remaining open questions.

\section{Theoretical background \label{sec:theory}}

In this work, we focus on the idealized setting of incompressible strong MHD turbulence driven at large scales in a periodic box. We further consider the limit of reduced MHD (RMHD), which describes the large-scale dynamics of plasmas with a uniform background magnetic field $\boldsymbol{B}_0 = B_0 \hat{\boldsymbol{z}}$ that is strong relative to turbulent fluctuations (i.e., $B_0 \gg b_\text{rms}$) and with typical gradients along $\boldsymbol{B}_0$ being much smaller than those perpendicular to $\boldsymbol{B}_0$. In this limit, the fluctuating components of the magnetic field and flow velocity along $\boldsymbol{B}_0$ are negligible, and the MHD equations can be written in terms of stream function $\phi$ and magnetic flux function $\psi$ as \citep{kadomtsev_kantorovich_1974, strauss_1976}
\begin{align}
\partial_t \psi+ \boldsymbol{v}\cdot\nabla\psi &= B_0 \partial_z \phi + \eta \nabla^2 \psi \nonumber \\
\partial_t \omega+ \boldsymbol{v} \cdot \nabla \omega - \boldsymbol{b} \cdot \nabla j &= B_0 \partial_z j+ \nu \nabla^2 \omega \, , \label{eqs:rmhd}
\end{align}
where $\boldsymbol{b} = \hat{\boldsymbol{z}} \times \nabla \psi$, $\boldsymbol{v} = \hat{\boldsymbol{z}} \times \nabla \phi$, $\omega = \nabla^2_\bot \phi$, and $j = \nabla^2_\bot \psi$. Note that $\omega$ and $j$ are the components of vorticity and current density in the direction of the guide field, respectively (with other components vanishing). We have rescaled magnetic field by $1/\sqrt{4 \pi \rho}$ so that the background field equals the Alfv\'{e}n velocity, $B_0 = v_A$. RMHD is generally valid at sufficiently small scales (relative to the driving scale) in turbulence with a critically-balanced energy cascade. Furthermore, RMHD is applicable to both collisional and collisionless plasmas as long as dynamical scales are larger than the characteristic microphysical scales (e.g., the ion gyroradius or ion skin depth) \citep{schekochihin_etal_2009}. Although resistivity $\eta$ and viscosity $\nu$ do not accurately describe the mechanisms of dissipation in many natural plasmas, they provide a physical energy sink which can be easily measured in simulations.

The foundations of inertial-range incompressible MHD turbulence were established in a landmark paper by Goldreich and Sridhar \citep{goldreich_sridhar_1995}, who proposed a phenomenological model for strong MHD turbulence based on critical balance, which posits that the timescale for linear Alfv\'{e}n waves traveling along the local background field, $\tau_A = l_\parallel/v_A$, equals the timescale for the nonlinear cascade in the transverse direction, $\tau_\text{NL} = l_\perp/\delta v_l$, where $l_\parallel$ and $l_\perp$ are the sizes of the eddy in the parallel and perpendicular direction, and $\delta v_l \sim l_\perp^{1/3}$ is the velocity increment at that scale. This phenomenology predicts scale-dependent anisotropy, an inertial-range energy spectrum of $E(k_\perp) \sim k_\perp^{-5/3}$ where $k_\perp$ is the wavevector perpendicular to the local background field, and dissipation scales given by $l_{\perp,\eta} \sim \eta^{3/4}$ and $l_{\parallel,\eta} \sim \eta^{1/2}$ (assuming $\eta = \nu$). The scale-dependent dynamic alignment between $\boldsymbol{v}$ and $\boldsymbol{b}$ modifies some of these scalings\citep{boldyrev_2005, boldyrev_2006}, making eddies anisotropic in three directions and producing a shallower energy spectrum $E(k_\perp) \sim k_\perp^{-3/2}$, with dissipation scales given by $l_{\perp,\eta} \sim \eta^{2/3}$ and $l_{\parallel,\eta} \sim \eta^{1/3}$.

Extending the phenomenology of inertial-range MHD turbulence to describe intermittent structures is nontrivial for a number of reasons. Firstly, most phenomenological theories ignore intermittency, which gives only a small correction to low-order statistics (e.g., the energy spectra). Secondly, the theories are strictly valid only in the inertial range and therefore do not address dynamics near the dissipation scale, where intermittency is most conspicuous. Thirdly, the theories tend to describe the typical scalings of space-filling eddies, whereas intermittency is manifest in thin structures that occupy a small fraction of space.

We note that, despite these issues, several phenomenological models for intermittency in MHD turbulence have been proposed \citep{grauer_etal_1994, politano_pouquet_1995, muller_biskamp_2000, chandran_etal_2015}. Motivated by the success of the She and Leveque model for describing intermittency in hydrodynamic turbulence \citep{she_leveque_1994}, these models assume log-Poisson statistics in order to predict the scaling exponents of structure functions, and are able to reproduce measurements in numerical simulations and in the solar wind reasonably well. However, their implications for the sizes, intensities, and morphology of dissipative structures are less clear. The recent model by Chandran et al. (2015) accounts for scale-dependent dynamic alignment and quantitatively predicts the 3D anisotropy of intense eddies, which can potentially be linked to intermittent dissipative structures \citep{chandran_etal_2015}.

As already mentioned, intermittency is most evident in the small-scale dynamics of turbulence, involving gradients of the magnetic and velocity fields. It is therefore natural to focus on the current density $j$, which is directly associated with the resistive energy dissipation rate per unit volume, $\eta j^2$, and the vorticity $\omega$, since $\nu \omega^2$ equals the viscous energy dissipation rate per unit volume when averaged over the system. For theoretical purposes, it is often more tractable to consider structures in the Els\"{a}sser vorticities, $\omega^\pm = \omega \pm j$. Structures in all of these quantities will be considered in this work.

There is no unique and ideal method for identifying intermittent structures \citep{kida_miura_1998, kolar_2007}. For simplicity, we define a structure to be a connected region in space in which the magnitude of a given intermittent field $f(\boldsymbol{x})$ exceeds a threshold parameter, $f_\text{thr}$, and is bounded by an isosurface at $|f| = f_\text{thr}$. Here, $f$ is any intermittent quantity such as $j$, $\omega$, and $\omega^\pm$. The only free methodological parameter is $f_\text{thr}$, which should be relatively large, e.g., $f_\text{thr} > f_\text{rms}$, so that structures occupy well-defined regions that do not span the entire system. Formally, this framework for identifying structures is mathematically well-defined and robustly characterizes the morphology of the field, including implicit information about higher-order correlations \citep{mecke_2000}. Potential drawbacks include the fact that the threshold parameter is arbitrary, overlapping structures cannot be individually distinguished, and the dynamics in the region below the threshold are not probed. In practice, the selected threshold can cause some structures to be under-resolved (if $f_\text{thr}$ is comparable to the local peak in $f$) and can affect whether two nearby peaks in $f$ are resolved into separate structures or an individual structure. However, in our experience, the statistical conclusions for well-resolved structures are insensitive to the threshold.

As a precursor to our numerical analysis, we first remark on some generic scaling properties of intermittent structures that may be anticipated from the RMHD equations. As we later show, these scalings are in agreement with our numerical simulations. To this end, we focus on Els\"{a}sser vorticity structures in an arbitrary turbulent field. Denoting the Els\"{a}sser potentials $\phi^\pm = \phi \pm \psi$ and Els\"{a}sser fields $\boldsymbol{z}^\pm = \boldsymbol{v} \pm \boldsymbol{b}$, the RMHD equations for Els\"{a}sser vorticities $\omega^\pm$ are
\begin{align}
\partial_t \omega^{\pm} + z^\mp_i \partial_i \omega^\pm + \epsilon_{i j 3} \partial_i \partial_k \phi^\mp \partial_j \partial_k \phi^\pm &= \pm B_0 \partial_3 \omega^\pm + \frac{\nu + \eta}{2} \partial_k \partial_k \omega^\pm + \frac{\nu - \eta}{2} \partial_k \partial_k \omega^\mp \, , \label{eq:elsvort}
\end{align}
where $\partial_i$ denotes the $i$th component of the gradient, $\epsilon_{ijk}$ is the Levi-Civita symbol, and indices are summed over components perpendicular to $\boldsymbol{B}_0$ (i.e., $i,j,k \in \{1,2\}$). For the remainder of the paper, we take $\eta = \nu$ for simplicity. Consider a structure represented as an isolated volume ${\mathcal V}$ bounded by an isosurface at $\omega^+ = \omega_\text{thr}$, enclosing points satisfying $\omega^+ > \omega_\text{thr} > 0$. Integration of Eq.~\ref{eq:elsvort} across ${\mathcal V}$ eliminates the two advective terms $z^-_i \partial_i \omega^+$ and $B_0 \partial_3 \omega^+$ (via the divergence theorem), which do not contribute to the growth or decay of the structure but may influence its morphology and motion. The remaining terms are
\begin{align}
\int_{\mathcal V} dV \left( \partial_t \omega^+ + \epsilon_{i j 3} \partial_i \partial_k \phi^- \partial_j \partial_k \phi^+ - \eta \partial_k \partial_k \omega^+ \right) &= 0 \, . \label{eq:elsvortint}
\end{align}
The first term here describes the overall growth or decay of the structure, and can be written $(d/dt) \int_{\mathcal V} dV (\omega^+ - \omega_\text{thr})$. The second term is the vortex stretching term, which describes the contribution to growth or decay from nonlinear interactions with the opposite Els\"{a}sser population. The third term describes the decay due to dissipation. An analogous equation for structures in $\omega^-$ is obtained by interchanging Els\"{a}sser populations in Eq.~\ref{eq:elsvortint}.

The net motion of the structure is governed by the advective terms in the RMHD equations, $z^\mp_i \partial_i \omega^\pm$ and $\pm B_0 \partial_3 \omega^\pm$. In particular, $\omega^+$ structures are advected counter to the guide field $\boldsymbol{B}_0$ at the Alfv\'{e}n velocity, whereas $\omega^-$ structures are advected in the direction of $\boldsymbol{B}_0$. Since structures in $j$ and $\omega$ are superpositions of $\omega^\pm$ structures, they have no net motion (on average) but may have Alfv\'{e}nic growth and decay associated with collisions between $\omega^\pm$ structures.

We make a rough estimate for the thickness of the structure as follows. Consider a structure that is instantaneously stationary, $\int_{\mathcal V} dV \partial_t \omega^+ = 0$. Then the nonlinear term balances the dissipation in Eq.~\ref{eq:elsvortint}. First, we assume that the dissipative term sets the thickness of structures, so that $\nabla_\perp^2 \omega^+ \sim \omega^+/T_c^2$, where $T_c$ is the characteristic thickness. We then assume that the structure is thin, so that locally the gradients are predominantly transverse to the structure; then to leading order, $\epsilon_{ij3} \partial_i \partial_k \phi^- \partial_j \partial_k \phi^+ \sim \partial_1 \partial_2 \phi^- \omega^+/2$. By balancing the nonlinear and dissipative terms, we get $T_c \sim \sqrt{2\eta/\langle\partial_1 \partial_2\phi^-\rangle}$, where the brackets here denote an average across the structure. If we suppose that $\langle\partial_1 \partial_2 \phi^-\rangle$ is comparable to the rms fluctuations in vorticity, $\langle \partial_1 \partial_2 \phi^- \rangle  \sim \omega^-_\text{rms} \sim \sqrt{\epsilon/\eta}$ where $\epsilon$ is the (global) mean energy dissipation rate (assuming balanced turbulence), then $T_c \sim \eta^{3/4}/{\epsilon^{1/4}}$, which is proportional to the Goldreich-Sridhar dissipation scale. If scale-dependent correlations between $\partial_1 \partial_2 \phi^-$ and $\omega^+$ are present, then the scaling may differ from this; in particular, an anti-correlation will cause a shallower scaling of $T_c$ with $\eta$. These considerations suggest that the thickness of the structure will lie near the dissipation scale.

Incidentally, if the vortex stretching term vanishes, e.g., by imposing $\phi^- \propto \phi^+$, then a thin structure with thickness near the dissipation scale will decay very rapidly. By estimating the dissipative term as $\nabla_\perp^2 \omega^+ \sim \omega^+/T_c^2$ and neglecting the nonlinear term in Eq.~\ref{eq:elsvortint}, we find
\begin{align}
\frac{d}{dt} \int_{\mathcal V} dV (\omega^+ - \omega_\text{thr}) \approx - \frac{\eta}{T_c^2} \int_{\mathcal V} dV (\omega^+ - \omega_\text{thr}) \, ,
\end{align}
so that 
\begin{align}
\int_{\mathcal V} dV (\omega^+ - \omega_\text{thr}) \propto \exp{(- t/\tau_\eta)} \, ,
\end{align}
with decay time $\tau_\eta = T_c^2/\eta$, which scales as $\tau_\eta \sim \eta^{1/2}/\epsilon^{1/2}$ for the above estimate of $T_c$. Hence, for a thin structure to survive for a substantial time, it must be fed by the nonlinearity.

Inside of the $\omega^+$ structure, one may expect the dynamical timescales associated with advection by $\boldsymbol{B}_0$ and large-scale fluctuations in $\boldsymbol{z}^-$ to be comparable,
\begin{align}
|\partial_t \omega^+| \sim |\boldsymbol{z}^- \cdot \nabla_\perp \omega^+| \sim |B_0 \partial_z \omega^+| \, , \end{align}
which leads to a characteristic timescale $\tau_c$ given by
\begin{align}
1/\tau_c \sim b_\text{rms}/W_c \sim B_0/L_c \, , \label{eq:balance}
\end{align}
where $L_c$ and $W_c$ are characteristic spatial scales parallel and perpendicular to the guide field; we assumed balanced turbulence so that $z^-_\text{rms} \sim b_\text{rms} \sim v_\text{rms}$ at large scales. Advection by large-scale fields therefore introduces two characteristic spatial scales which are proportional to each other, $L_c \sim (B_0/b_\text{rms}) W_c$, and a timescale given by $\tau_c \sim L_c/B_0$. The quantities $L_c$, $W_c$, and $\tau_c$ are naturally associated with the length, width, and lifetime of the structures, respectively. As we show in the numerical results that follow, these quantities are distributed across inertial-range scales, in contrast to dissipation-scale thickness $T_c$.

\section{Numerical results}

\subsection{Simulations}

We now describe the statistical analysis of intermittent structures in our numerical simulations of MHD turbulence, mainly focusing on current sheets. Turbulence is driven at large scales in a periodic box; the energy then cascades through an inertial range until it is lost in the dissipation range. Additional details on our pseudo-spectral simulations are described elsewhere\citep{perez_boldyrev_2010, perez_etal_2012}. The RMHD equations are solved in a box that is elongated along the guide field by a factor of $L_\parallel/L_\perp = 6$ (where $L_\perp = 2 \pi$ in simulation units), to compensate for the anisotropy due to the strong guide field, which is fixed at  $B_0/b_\text{rms} \approx 5$. The turbulence is driven by colliding Alfv\'en modes, excited by statistically independent random forces in Fourier space at low wave-numbers $2\pi/L_{\perp} \leq k_{x,y} \leq 2 (2\pi/L_{\perp})$, $k_z = 2\pi/L_\parallel$. The Fourier coefficients of the forcing in this range are Gaussian random numbers with amplitudes chosen so that $b_\text{rms} \sim v_\text{rms}\sim 1$. The random values of the different Fourier components of the forces are refreshed independently on average about $10$ times per eddy turnover time. The Reynolds number is given by $\operatorname{Re} = v_\text{rms} (L_\perp/2\pi) / \nu$, with magnetic Prandtl number $\operatorname{Pm} = \nu/\eta = 1$. The analysis is performed for 15 snapshots (spaced at intervals of one eddy-turnover time during statistical steady state) for independent runs with $Re = 1000$, $Re = 1800$, and $Re = 3200$ on $1024^3$ lattices, and also for 9 snapshots with $Re = 9000$ on a $2048^3$ lattice; these runs were also analyzed in our previous work \citep{zhdankin_etal_2014}. Unless otherwise noted, we show results from the $Re = 9000$ case.

Since $\operatorname{Pm} = 1$ and turbulence is balanced ($\omega^+_\text{rms} \approx \omega^-_\text{rms}$) in the simulations, the total energy dissipation rate is ${\mathcal E}_\text{tot} = V_\text{tot} \eta ( j_\text{rms}^2 + \omega_\text{rms}^2) = V_\text{tot} \eta (\omega^+_\text{rms})^2$ where $V_\text{tot}$ is the system volume. Since ${\mathcal E}_\text{tot}$ is fixed (balancing the energy input from driving), we can anticipate the rms fluctuations to scale as $f_\text{rms} \propto \sqrt{{\mathcal E}_\text{tot}/\eta V_\text{tot}} \propto \operatorname{Re}^{1/2}$ for $f \in \{j,\omega,\omega^+\}$. We therefore take the threshold values relative to the rms value when comparing intermittency in simulations with varying $\operatorname{Re}$; we denote $\tilde{f} = f / f_\text{rms}$.

\subsection{Cumulative distributions}

\begin{figure}
\centering
\includegraphics[width=0.7\columnwidth]{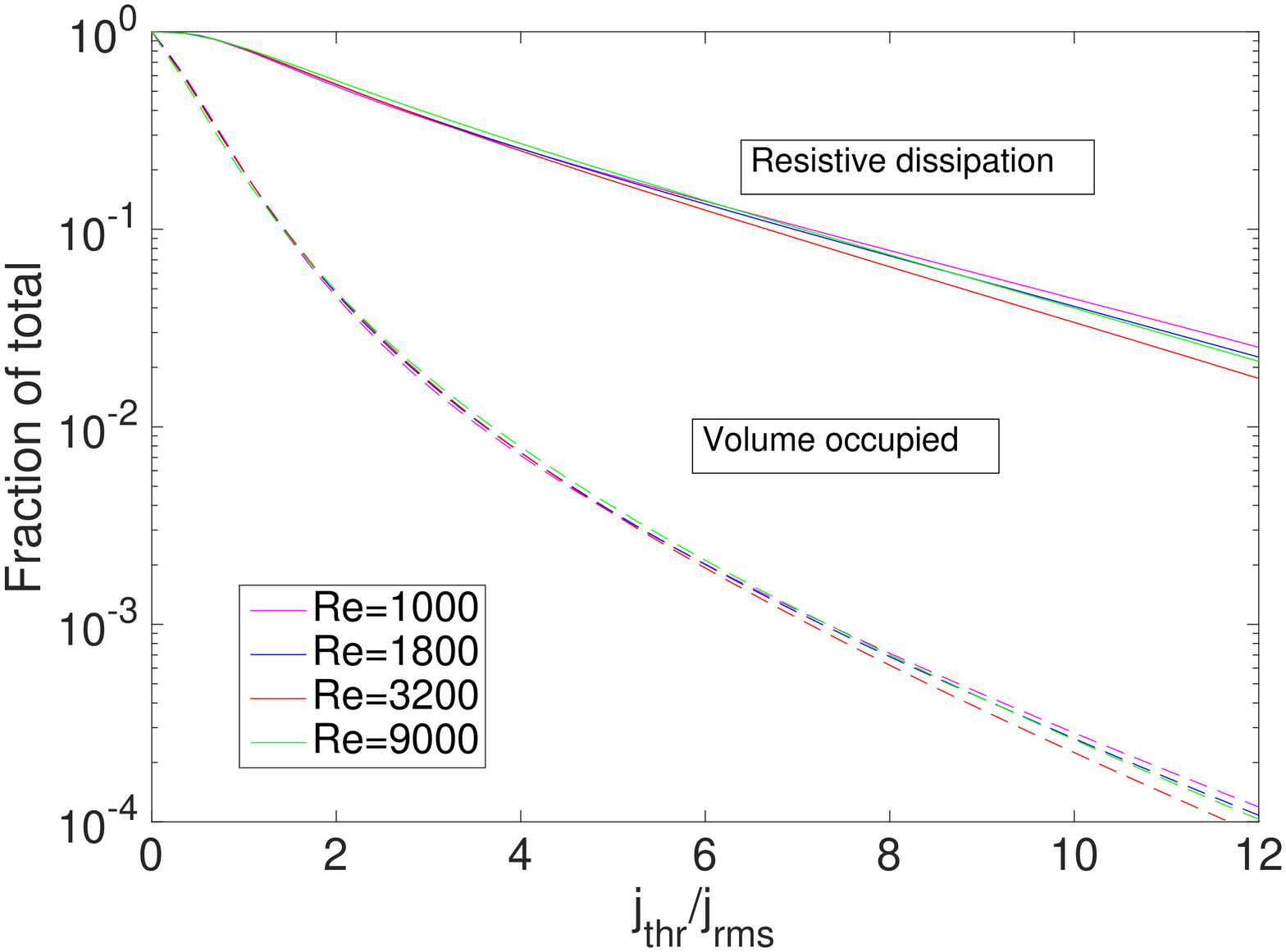}
 \includegraphics[width=0.7\columnwidth]{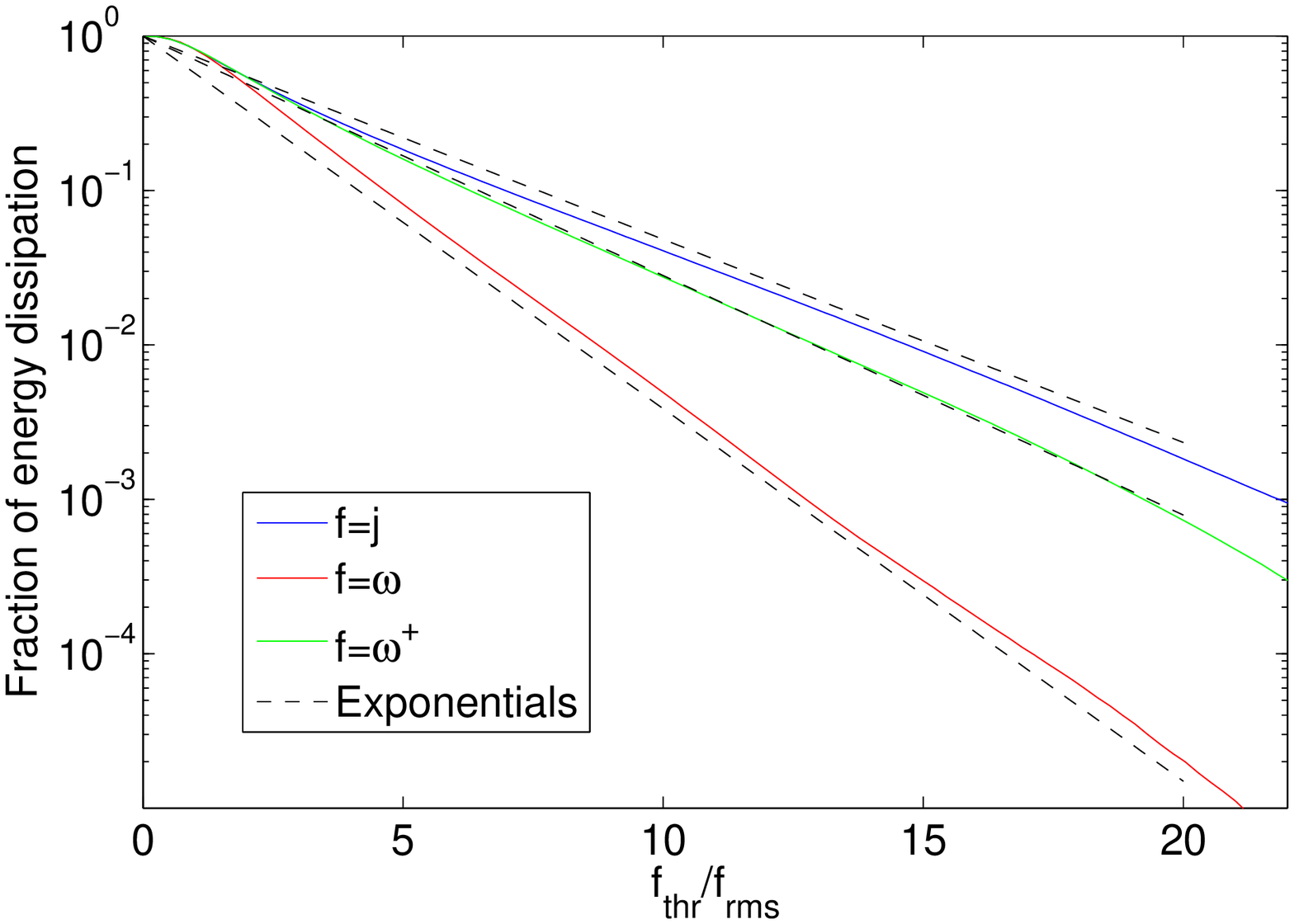}
 \caption{Top panel: the fraction of overall resistive energy dissipation ${\mathcal E}^j_\text{cum}(\tilde{j}_\text{thr})/{\mathcal E}^j_\text{tot}$ (solid lines) and fraction of volume occupied $V^j_\text{cum}(\tilde{j}_\text{thr})/V_\text{tot}$ (dashed lines) by structures with current densities $|j|/j_\text{rms} > \tilde{j}_\text{thr}$. The colors correspond to $\operatorname{Re} = 1000$ (magenta), $\operatorname{Re} = 1800$ (blue), $\operatorname{Re} = 3200$ (red), and $\operatorname{Re} = 9000$ (green). Bottom panel: ${\mathcal E}^j_\text{cum}(\tilde{j}_\text{thr})/{\mathcal E}^j_\text{tot}$ (blue) compared to the viscous dissipation in vorticity structures ${\mathcal E}^\omega_\text{cum}(\tilde{\omega}_\text{thr})/{\mathcal E}^\omega_\text{tot}$ (red) and dissipation in Els\"{a}sser vorticity structures ${\mathcal E}^{\omega^+}_\text{cum}(\tilde{\omega}^+_\text{thr})/{\mathcal E}^{\omega^+}_\text{tot}$ (green). Exponential fits $\exp{(-x/x_0)}$ with $x = f_\text{thr}/f_\text{rms}$ and $x_0 \in \{3.3,2.8,1.8\}$ for $f \in \{j,\omega^+,\omega\}$ are also shown (black dashed lines). \label{fig:energy_volume}}
 \end{figure}

The presence of intermittency in the dissipative field $f \in \{j,\omega,\omega^+\}$ can be inferred, to some extent, from the cumulative energy dissipation rate ${\mathcal E}^f_\text{cum}(\tilde{f}_\text{thr})$ and cumulative volume $V^f_\text{cum}(\tilde{f}_\text{thr})$ conditioned on the normalized threshold $\tilde{f}_\text{thr}$. In terms of the probability density function for absolute value of $\tilde{f}$, $P(|\tilde{f}|)$, these are given by
\begin{align}
{\mathcal E}^f_\text{cum}(\tilde{f}_\text{thr})/{\mathcal E}^f_\text{tot} &= \int_{\tilde{f}_\text{thr}}^\infty d|\tilde{f}| P(|\tilde{f}|) |\tilde{f}|^2 \, , \nonumber \\
V^f_\text{cum}(\tilde{f}_\text{thr})/V_\text{tot} &= \int_{\tilde{f}_\text{thr}}^\infty d|\tilde{f}| P(|\tilde{f}|) \, ,
\end{align}
where ${\mathcal E}^f_\text{tot} = V_\text{tot} \eta f_\text{rms}^2$ is the total energy dissipation rate in the system associated with $f$. In essence, these quantities represent the energy dissipated and volume occupied by structures in the field $f$ at the given threshold; specifically, ${\mathcal E}^j_\text{cum}$ is the resistive dissipation rate in current density structures, ${\mathcal E}^\omega_\text{cum}$ is the viscous dissipation rate in vorticity structures, and ${\mathcal E}^{\omega^+}_\text{cum}$ is the dissipation rate in the Els\"{a}sser vorticity structures.

We first consider cumulative distributions for the current density. As shown in Fig.~\ref{fig:energy_volume}, the fraction of total resistive dissipation ${\mathcal E}_\text{cum}^j(\tilde{j}_\text{thr})/{\mathcal E}_\text{tot}^j$ and fraction of volume occupied $V^j_\text{cum}(\tilde{j}_\text{thr})/V_\text{tot}$ are remarkably insensitive to $\operatorname{Re}$, with the former significantly exceeding the latter at high thresholds. For example, $50\%$ of the resistive energy dissipation occurs in regions with current densities exceeding $\tilde{j}_\text{thr} \approx 2.2$, at which current sheets are still visibly well-defined and occupy only about $3\%$ of the volume. On this basis, one can conclude that the majority of (resistive) energy dissipation occurs in intermittent structures. Similar cumulative distributions were investigated for numerical simulations of collisionless plasmas \citep{wan_etal_2012b, makwana_etal_2015} and line-tied MHD \citep{wan_etal_2014}, the latter of which is reported to be more intermittent than our case, with $50\%$ of dissipation occurring in $0.4\%$ of the volume.

To a first approximation, the tail of ${\mathcal E}^j_\text{cum}(\tilde{j}_\text{thr})/{\mathcal E}^j_\text{tot}$ can be fit remarkably well by an exponential, $\exp{(-\tilde{j}_\text{thr}/3.3)}$, as shown in the second panel of Fig.~\ref{fig:energy_volume}. For comparison, we also show the viscous dissipation in vorticity structures, ${\mathcal E}^\omega_\text{cum}(\tilde{\omega}_\text{thr})/{\mathcal E}^\omega_\text{tot}$, which decays more steeply with threshold $\tilde{\omega}_\text{thr}$ than the resistive case, roughly as $\exp{(-\tilde{\omega}_\text{thr}/1.8)}$, implying that $\omega$ is less intermittent than $j$. We also show the dissipation in Els\"{a}sser vorticity structures, ${\mathcal E}^{\omega^+}_\text{cum}(\tilde{\omega}^+_\text{thr})/{\mathcal E}_\text{tot}$, which is fit by $\exp{(-\tilde{\omega}^+_\text{thr}/2.8)}$.

\begin{figure}
\centering
\includegraphics[width=0.75\columnwidth]{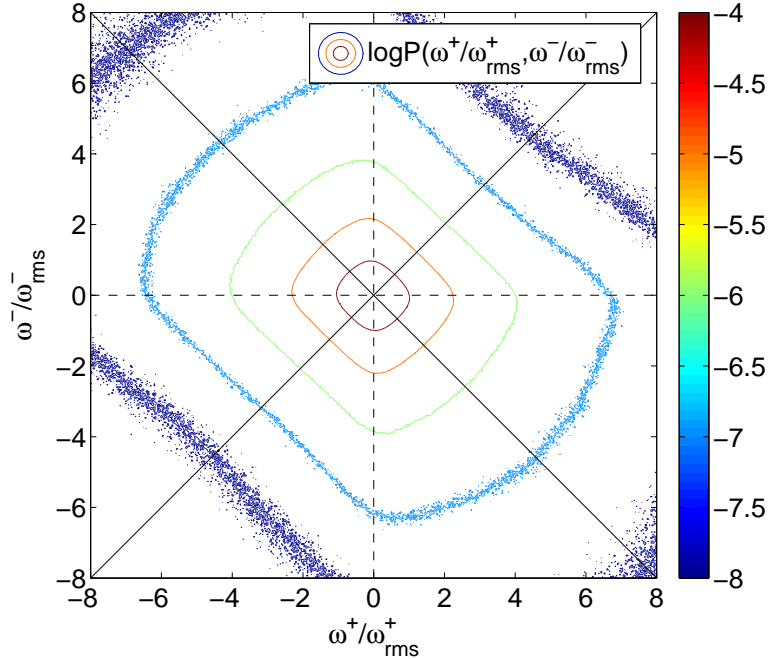}
 \caption{The 2D probability density function $P(\tilde{\omega}^+, \tilde{\omega}^-)$. Dashed black lines represent the axes for $\omega^-$ and $\omega^+$, while solid black lines represent the axes for $j$ (going from top-left to bottom-right) and $\omega$ (going from bottom-left to top-right). \label{fig:corr}}
 \end{figure}

The correlations between the $j$, $\omega$, and $\omega^\pm$ can be ascertained from the 2D probability density function for Els\"{a}sser vorticities, $P(\omega^+, \omega^-)$, which is equivalent to $P(j,\omega)$ rotated clockwise by 45 degrees. We show the contours of $P(\tilde{\omega}^+, \tilde{\omega}^-)$ for $\operatorname{Re} = 3200$ in Fig.~\ref{fig:corr}. This distribution is symmetric across the (diagonal) $\omega$ and $j$ axes, as required by the symmetries of the RMHD equations, but is not symmetric across the $\omega^-$ and $\omega^+$ axes. Instead, $\omega^-$ and $\omega^+$ have a small tendency to have opposite signs rather than equal signs. This asymmetry can be inferred from the RMHD equations: the vortex stretching term  $\epsilon_{i j 3} \partial_i \partial_k \phi^- \partial_j \partial_k \phi^+$ acts with opposite signs on the two Els\"{a}sser populations, locally skewing $\omega^+$ and $\omega^-$ toward opposite signs. As a consequence, large values of $j$ are more likely than large values of $\omega$, implying that $j_\text{rms} > \omega_\text{rms}$ and that the resistive dissipation will generally be larger than the viscous dissipation (despite $\operatorname{Pm} = 1$). Indeed, we find that the ratio of resistive-to-viscous dissipation $\mathcal{E}^\eta_\text{tot}/\mathcal{E}^\nu_\text{tot}$ in our simulations varies from approximately $1.67$ at $\operatorname{Re} = 1000$ to $1.42$ at $\operatorname{Re} = 9000$, indicating a mismatch between the two types of dissipation. A similar mismatch, which varies with $\operatorname{Pm}$, has been noted in previous studies of flow-driven MHD turbulence with no guide field \citep{sahoo_etal_2011, brandenburg_2014}.

\subsection{Statistical analysis}

We next consider the statistical properties of dissipative structures in the intermittent fields $f \in \{j,\omega,\omega^+\}$. Each structure is represented as a set of spatially-connected points satisfying $|f| > f_\text{thr}$, where two points on the lattice are spatially-connected if they are separated by strictly less than $2$ lattice spacings. Each structure in $\{j,\omega,\omega^+\}$ has a corresponding energy dissipation rate given by ${\mathcal E} = \int dV \{ \eta j^2, \nu \omega^2, \eta (\omega^+)^2/2 \}$, where integration is performed across the volume of the individual structure. For the following analysis, we take a threshold of $f_\text{thr}/f_\text{rms} \approx 3.75$, which captures the most intense and well-defined structures while being low enough to give a large sample of structures (typically thousands of well-resolved structures per snapshot). At this threshold, structures occupy roughly $1\%$ of the volume but have a large contribution to the energy dissipation: current sheets contribute about $30\%$ of the overall resistive dissipation, vorticity sheets about $22\%$ of the viscous dissipation, and Els\"{a}sser vorticity structures about $36\%$ of the dissipation.

\begin{figure*}
\centering
\includegraphics[width=0.75\columnwidth]{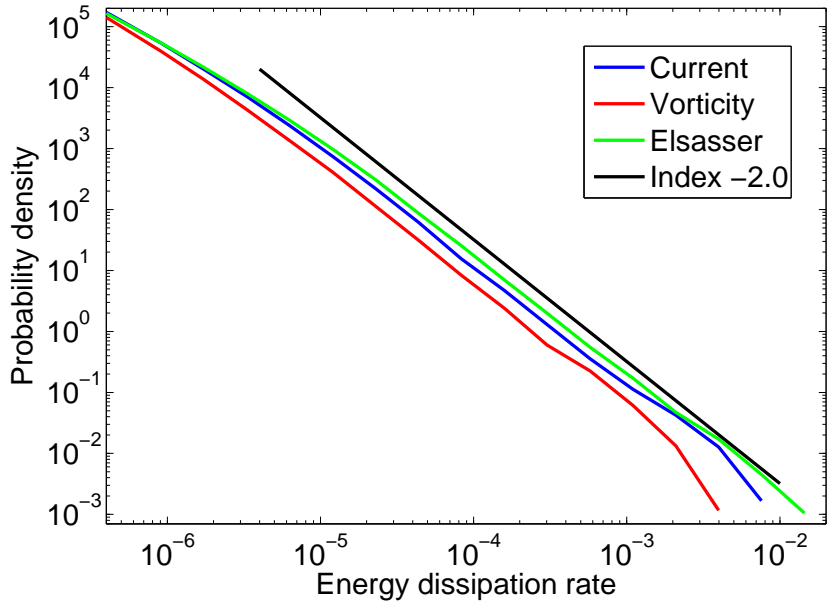}
 \caption{Probability distribution for energy dissipation rate ${\mathcal E}$ associated with structures in current density (blue), vorticity (red), and Els\"{a}sser vorticity (green). For comparison, a power law with the critical index of $-2.0$ is shown in black. \label{fig:diss}}
 \end{figure*}

We first show the probability distribution for the energy dissipation rates ${\mathcal E}$ associated with structures in $\{j,\omega,\omega^+\}$ in Fig.~\ref{fig:diss}. We find that this distribution has a power law with index very close to $-2.0$ for all three types of structures. As noted in our previous work\citep{zhdankin_etal_2014}, this index is a critical value in which both the weak structures and the strong structures contribute equally to the overall energy dissipation rate. It is remarkable that this critical index appears to robustly describe all three types of dissipative structures, suggesting that turbulence spreads dissipation across all available dissipation channels and energy scales.

Ignoring the finer features, each structure can be characterized by three scales. These are the length $L$, width $W$, and thickness $T$, with $L \ge W \ge T$. In previous works, we applied three distinct methods for measuring these spatial scales for current sheets; in this work, we use the Euclidean method from Zhdankin et al. 2014 \citep{zhdankin_etal_2013, zhdankin_etal_2014}. These scales give direct measurements of the size across certain parts of the structure, although they may not capture irregular morphologies very well. For length $L$, we take the maximum distance between any two points in the structure. For width $W$, we consider the plane orthogonal to the length and coinciding with the point of peak amplitude. We then take the maximum distance between any two points of the structure in this plane to be the width. The direction for thickness $T$ is then set to be orthogonal to length and width. We take $T$ to be the distance across the structure in this direction through the point of peak amplitude. All of these scales are measured in units of the perpendicular box size $L_\perp$.

We now consider the probability distributions for the characteristic scales of the intermittent structures. The probability distributions for length $L$, width $W$, and thickness $T$ of structures in $\{j,\omega,\omega^+\}$ are shown in Fig.~\ref{fig:diss}. For all three types of structures, we find that $L$ and $W$ both have robust power-law distributions with indices near $-3.3$ for scales spanning the inertial range. On the other hand, the distribution for $T$ decreases very rapidly at scales within the dissipation range, which implies that there are few intense structures with large thicknesses, although some dissipation may still occur inside weaker structures at those scales. As is now clear, intermittent structures in the current density, vorticity, and Els\"{a}sser vorticities all have nearly identical statistical properties.

\begin{figure*}
\centering
\includegraphics[width=0.32\columnwidth]{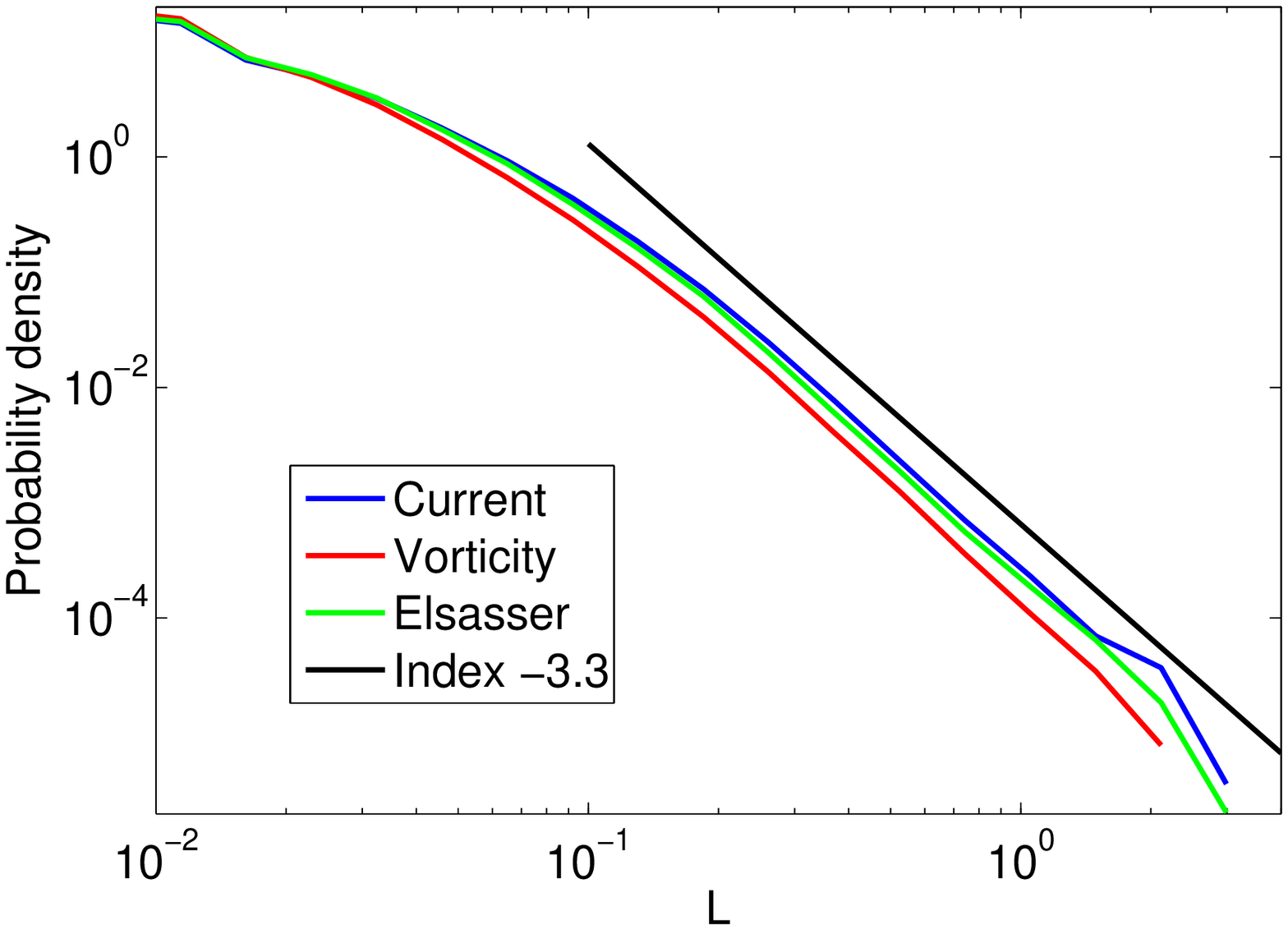}
\includegraphics[width=0.32\columnwidth]{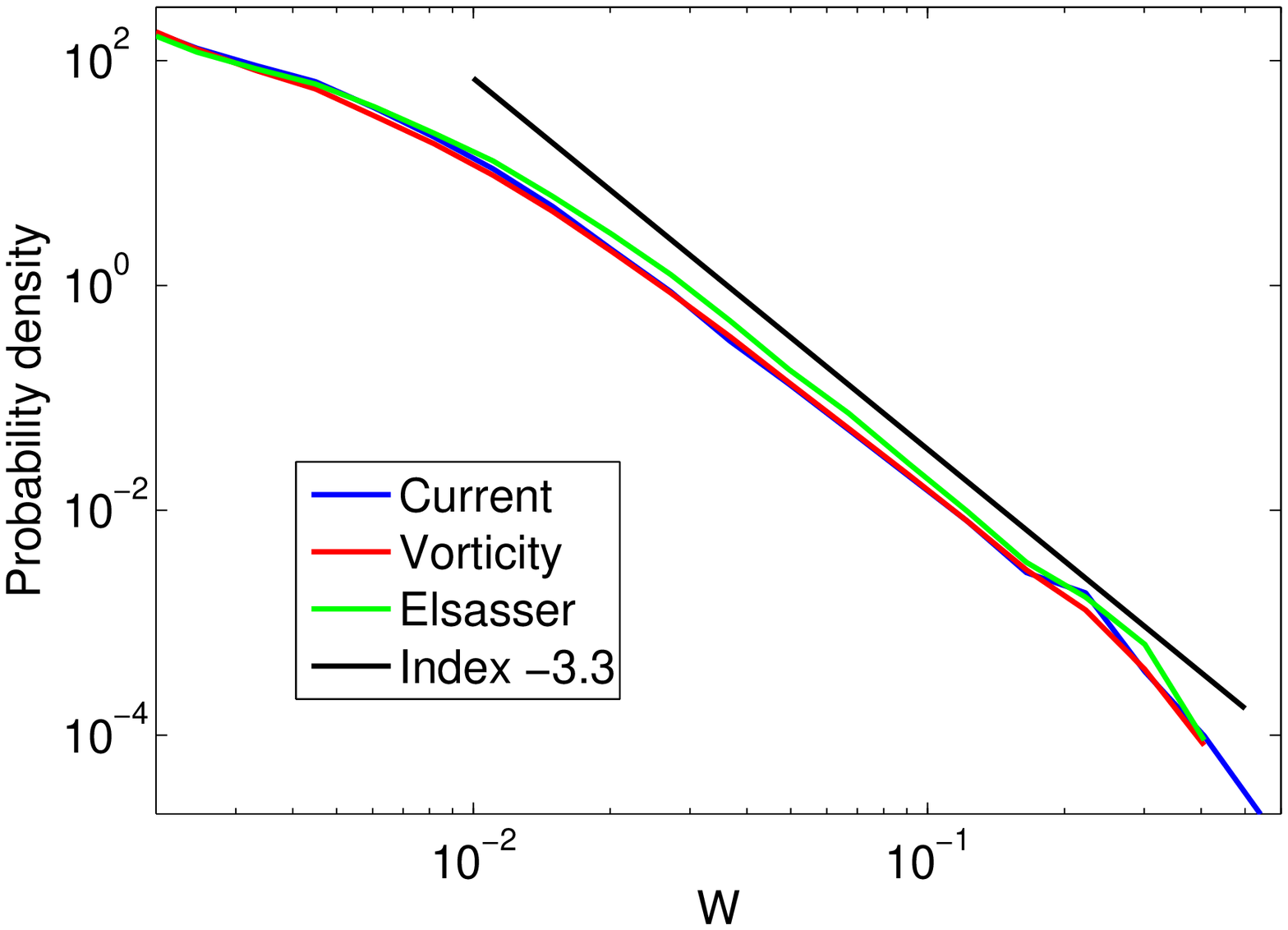}
\includegraphics[width=0.32\columnwidth]{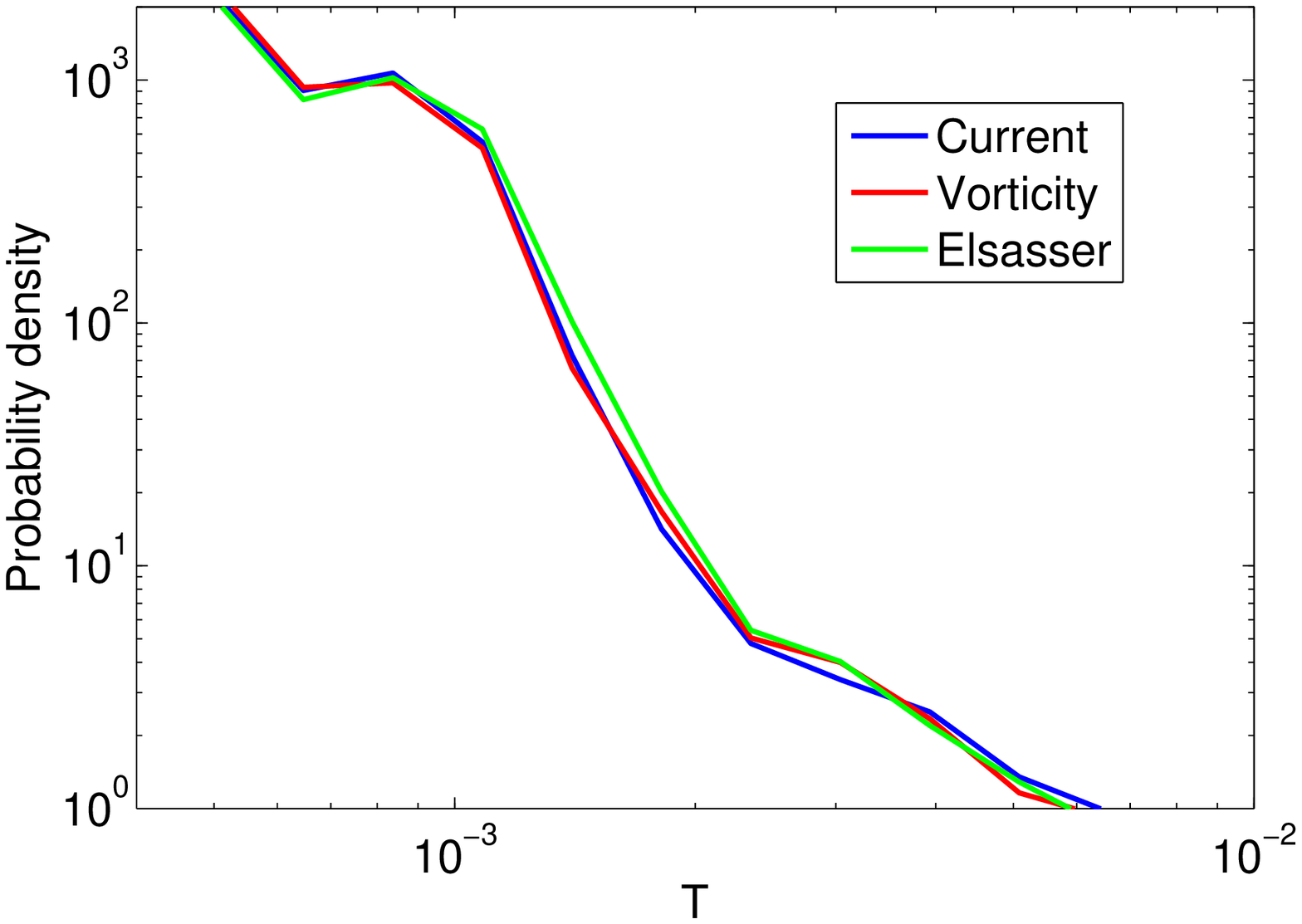}
 \caption{Probability distributions for length $L$, width $W$, and thickness $T$ of structures in current density (blue), vorticity (red), and Els\"{a}sser vorticity (green). For comparison, a power law with index $-3.3$ is shown (in black). \label{fig:all}}
 \end{figure*}

\begin{figure*}
\centering
 \includegraphics[width=0.32\columnwidth]{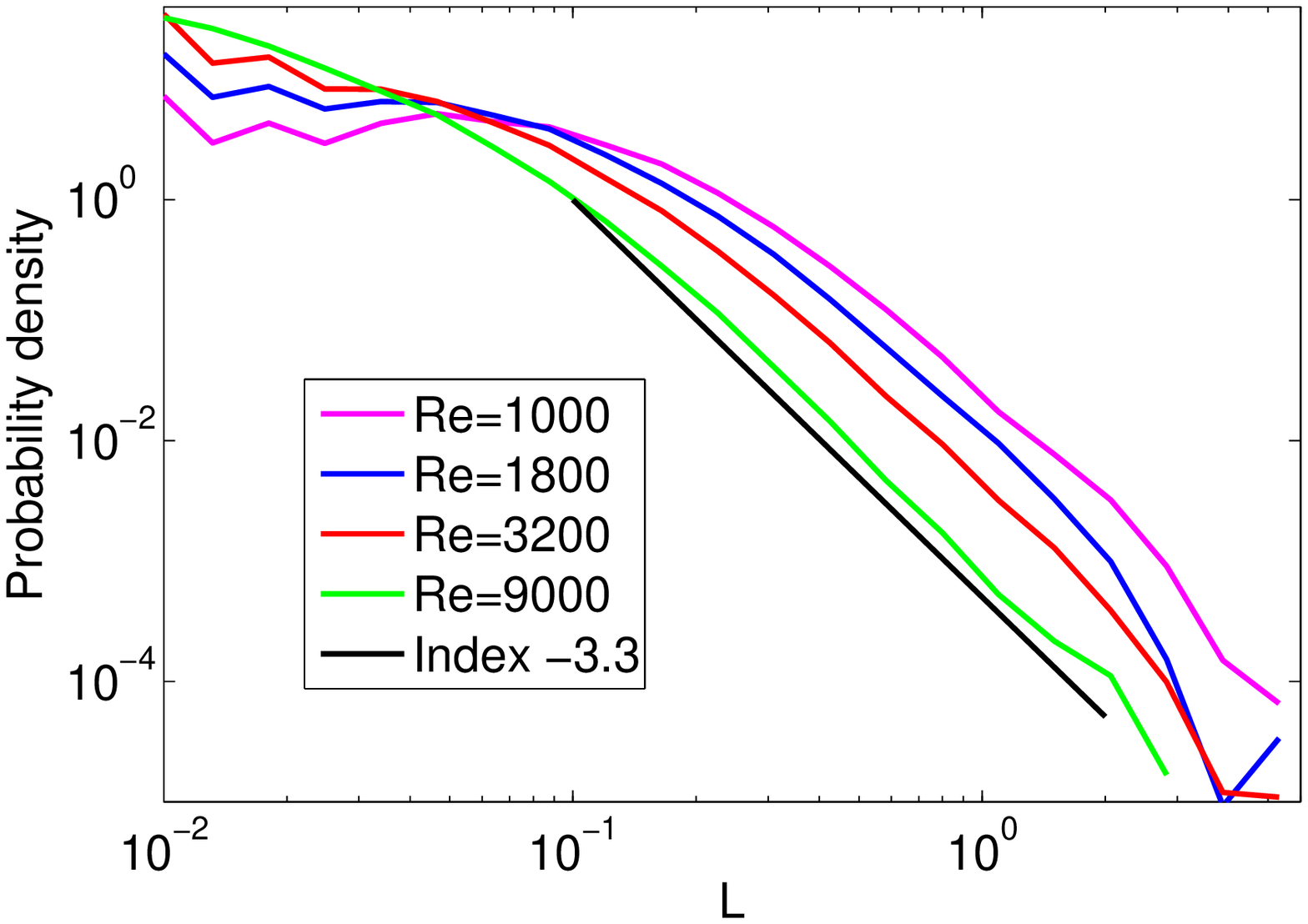}
 \includegraphics[width=0.32\columnwidth]{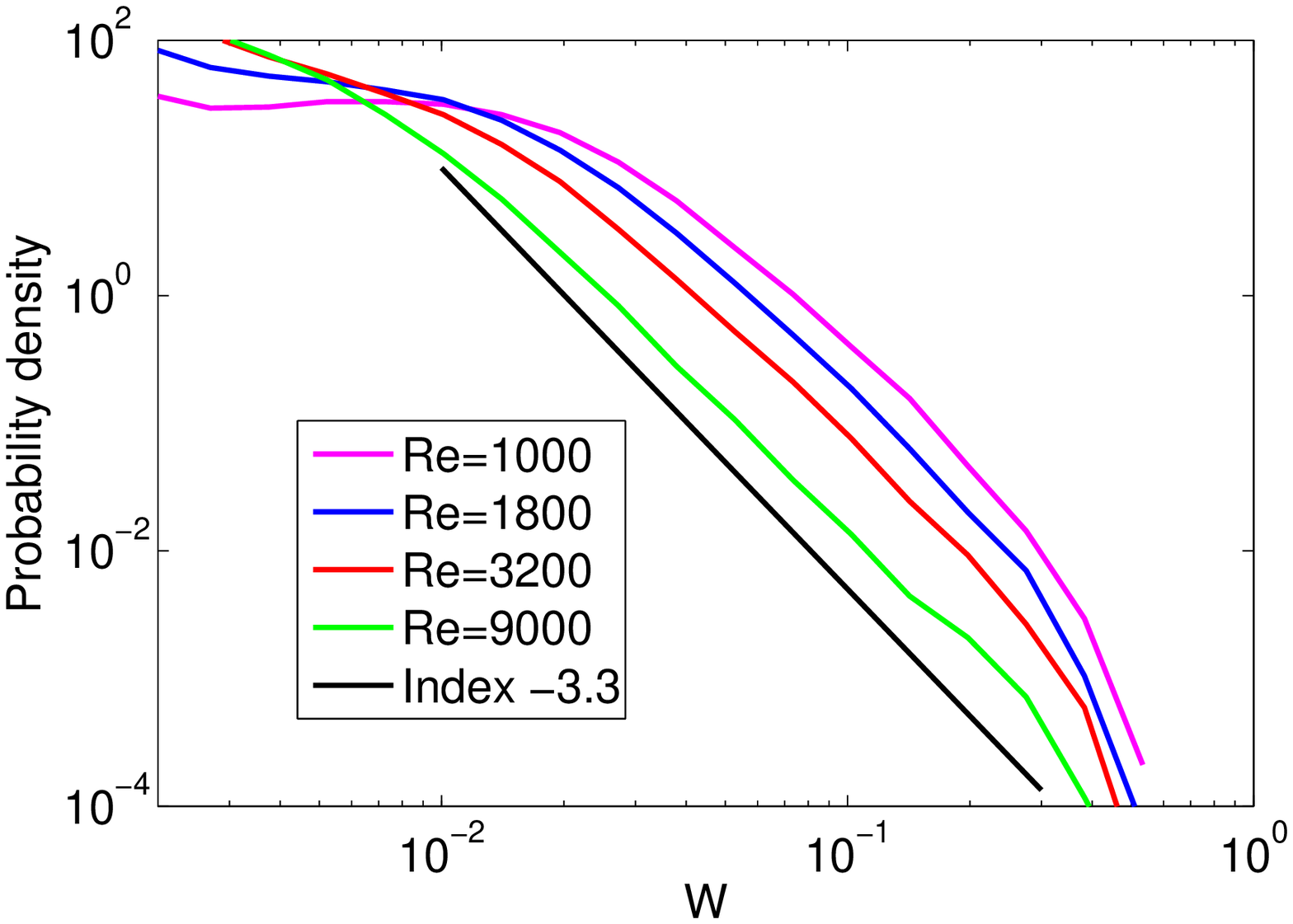}
 \includegraphics[width=0.32\columnwidth]{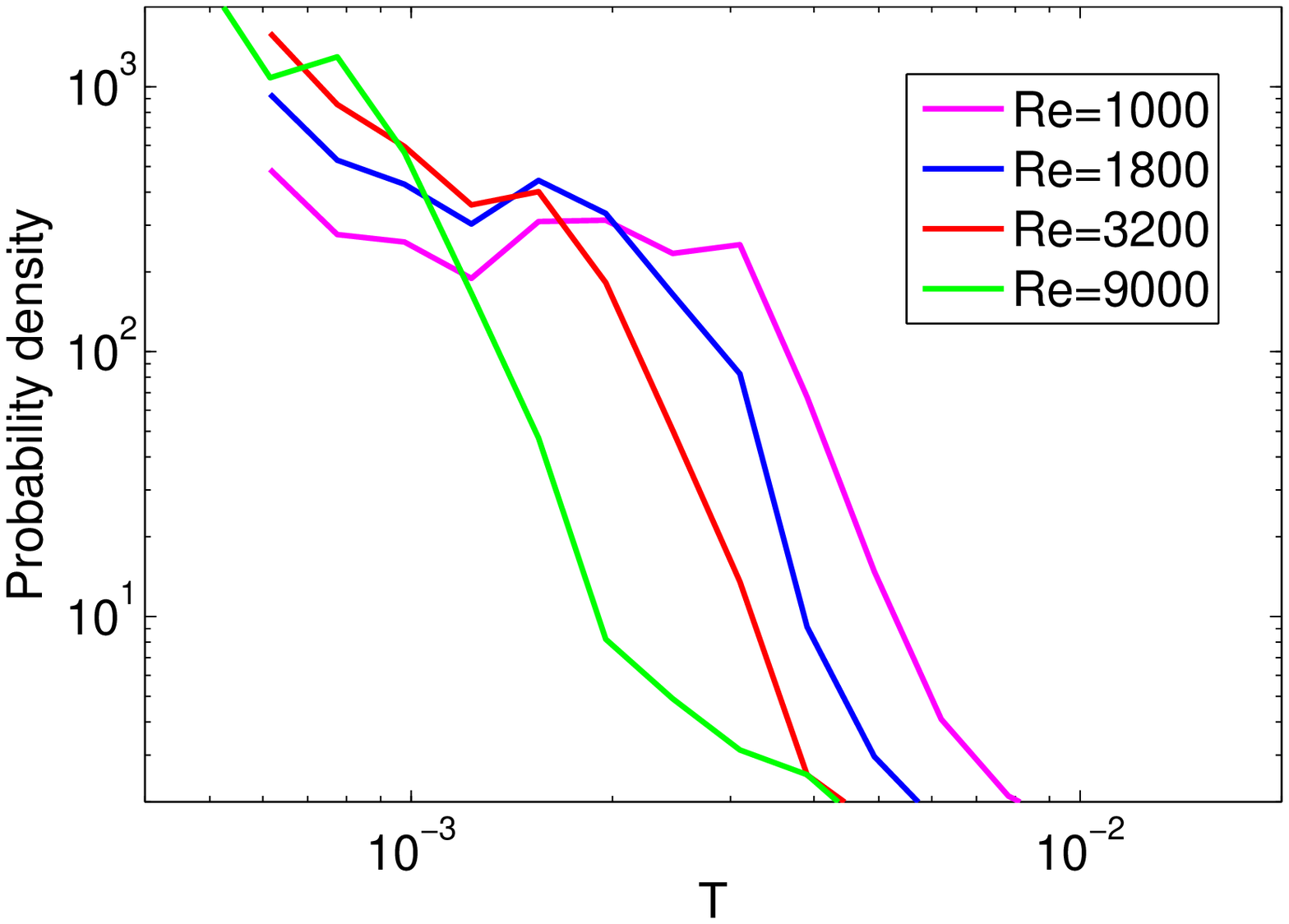}
\includegraphics[width=0.32\columnwidth]{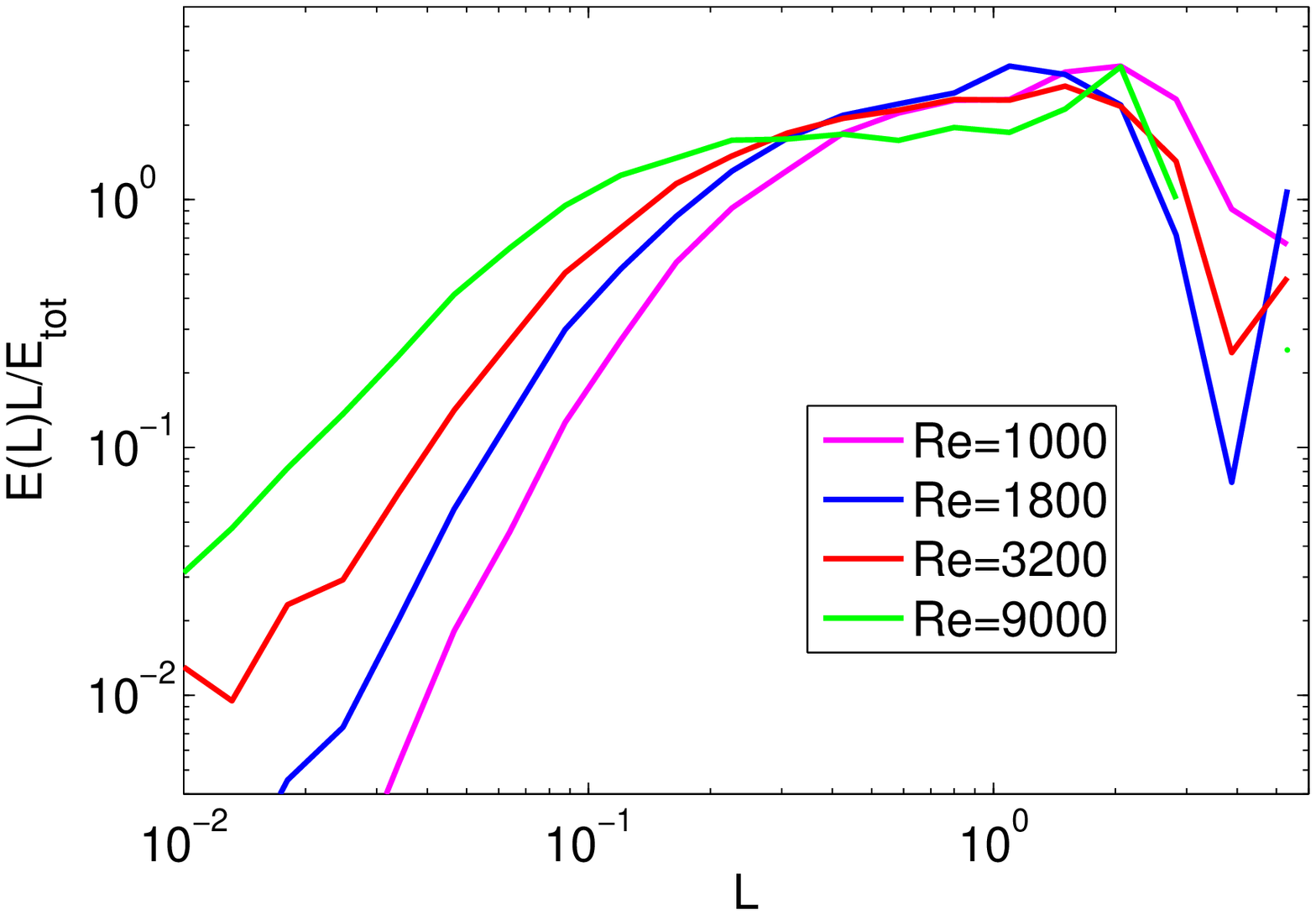}
 \includegraphics[width=0.32\columnwidth]{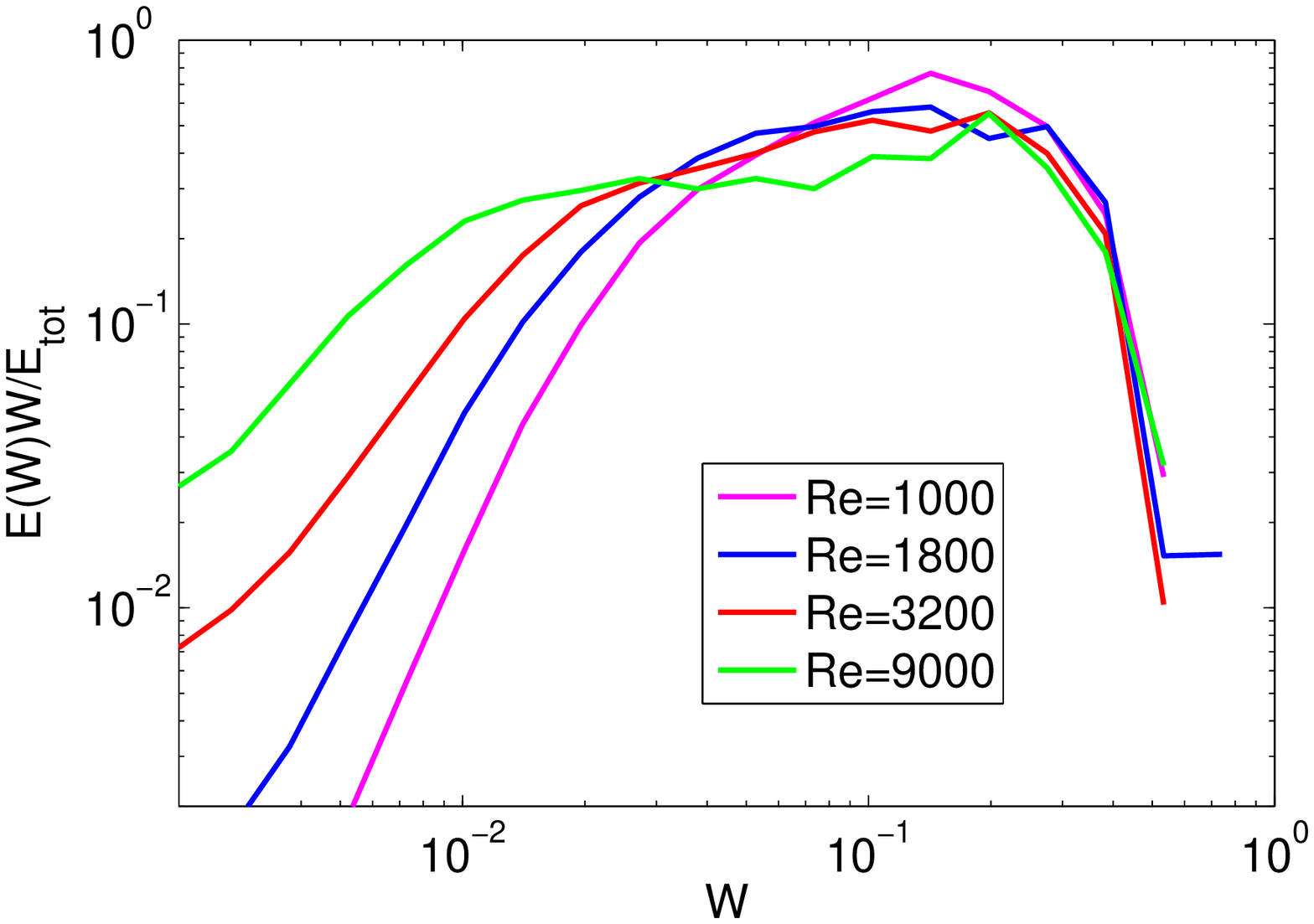}
 \includegraphics[width=0.32\columnwidth]{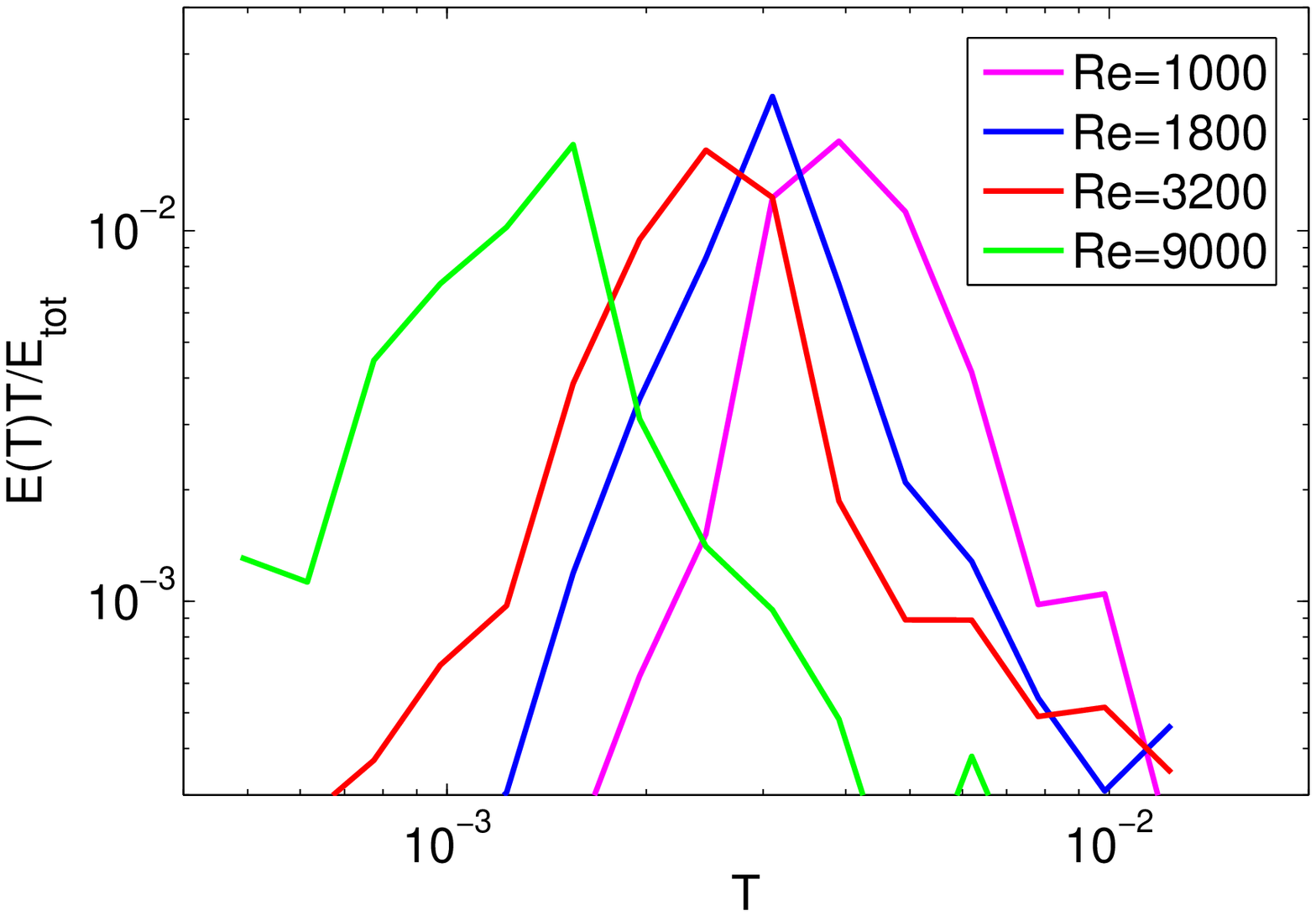}
\includegraphics[width=0.32\columnwidth]{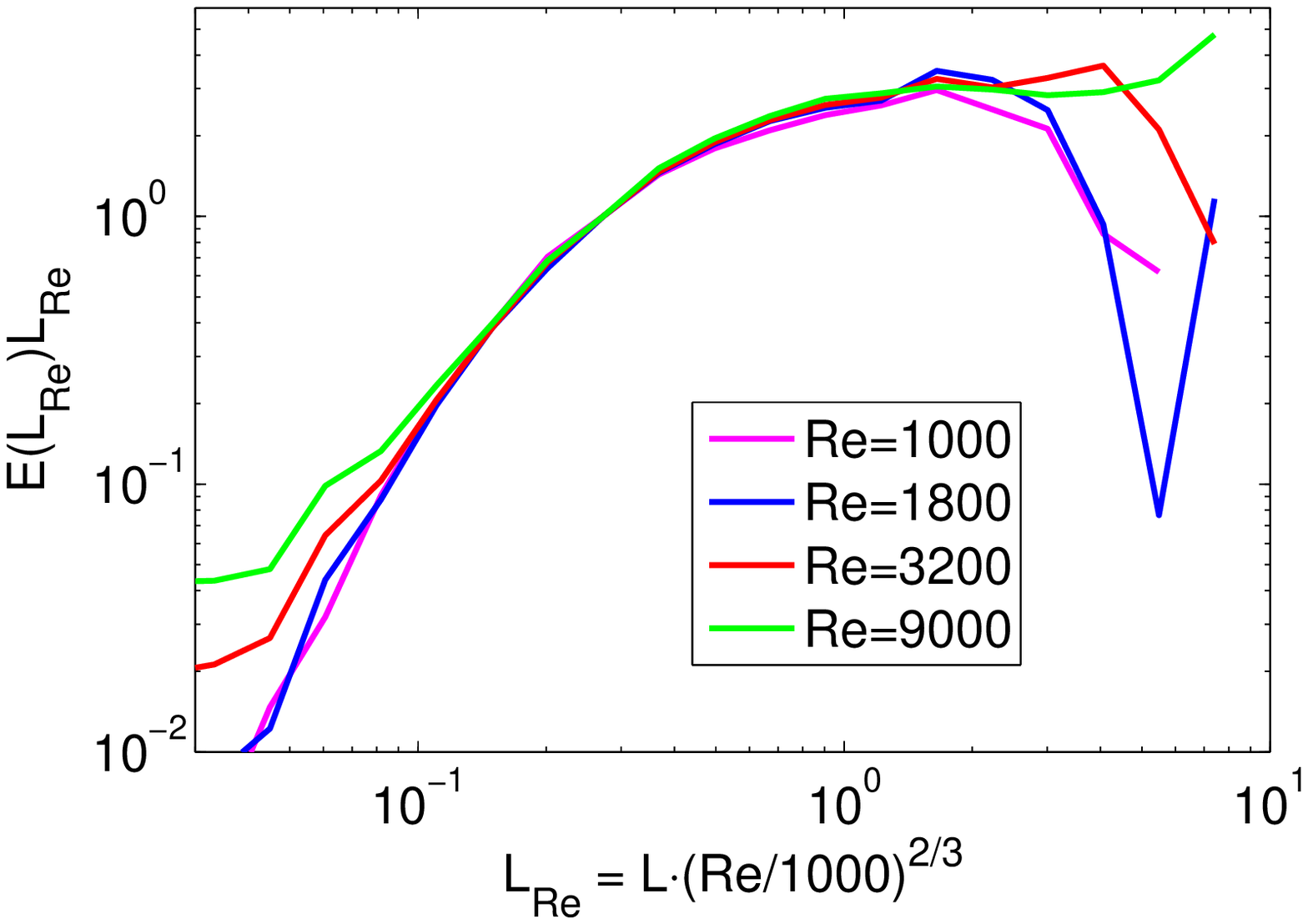}
 \includegraphics[width=0.32\columnwidth]{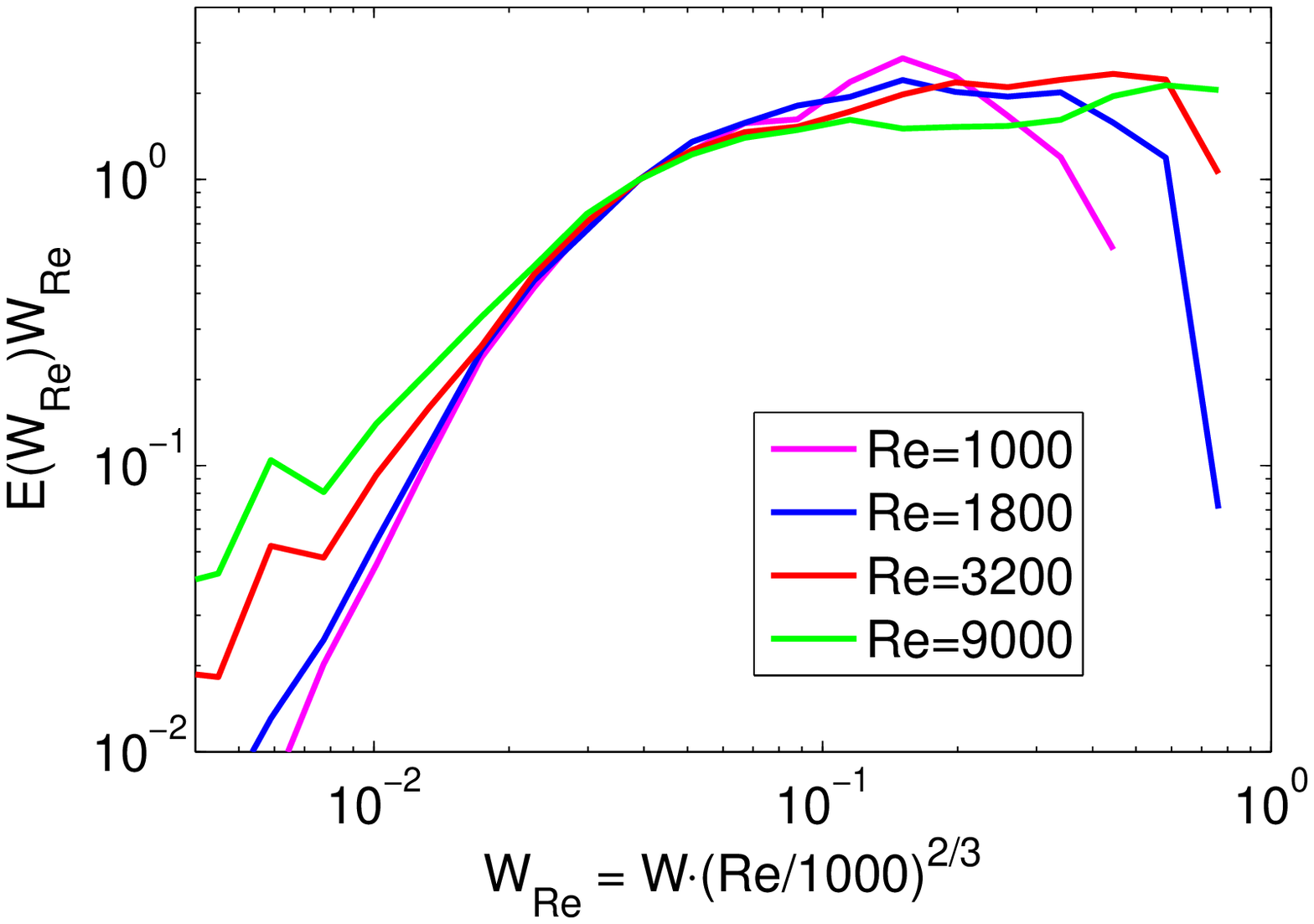}
 \includegraphics[width=0.32\columnwidth]{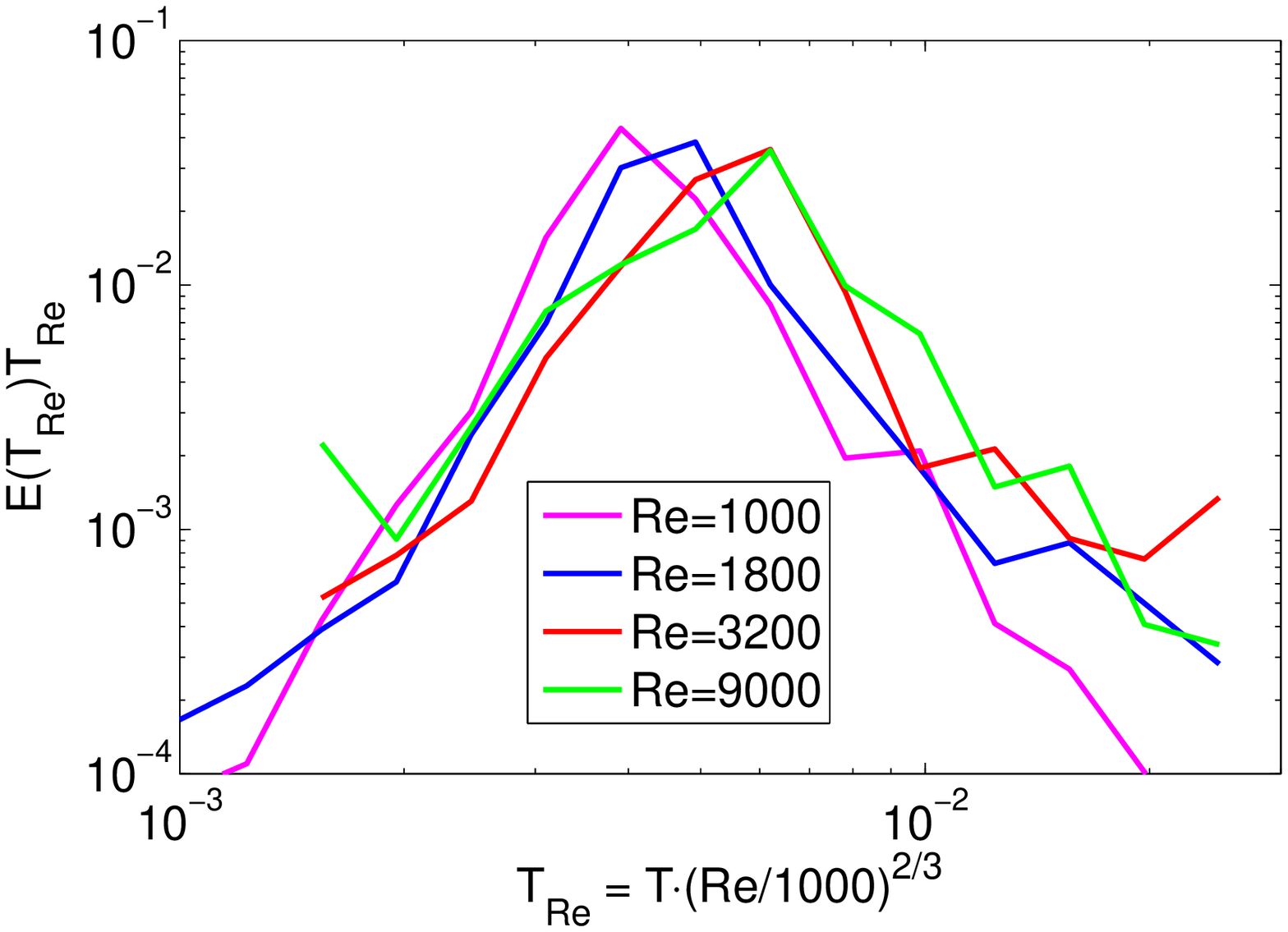}
 \caption{Three representations of the probability distributions for current sheet lengths $L$ (left column), widths $W$ (center column), and thicknesses $T$ (right column), for $\operatorname{Re} = 1000$ (magenta), $\operatorname{Re} = 1800$ (blue), $\operatorname{Re} = 3200$ (red), and $\operatorname{Re} = 9000$ (green). The top row shows the unweighted probability distributions, the center row shows the (compensated) dissipation-weighted distributions, and the bottom row shows the dissipation-weighted distributions for the quantities rescaled by a factor of $(\operatorname{Re}/1000)^{2/3}$ (with arbitrary normalization for the last case). \label{fig:dists}}
 \end{figure*}

We now consider the statistics of the structures in the current density in more detail (the remaining results are similar for vorticity and Els\"{a}sser vorticity structures). In the first row of Fig.~\ref{fig:dists}, the probability distributions for the spatial scales are shown for varying $\operatorname{Re}$. To determine the sizes of the most energetic structures, it is more transparent to consider the dissipation-weighted distributions $E(X)$\cite{zhdankin_etal_2014}. We define $E(X) dX$ to be the combined energy dissipation rate for structures with the measured scale between $X$ and $X + dX$. The maximum of the compensated dissipation-weighted distribution, $E(X)X$, indicates the scale of the structures which give the dominant contribution to the overall energy dissipation rate. We show $E(L)L$, $E(W)W$ and $E(T)T$ in Fig.~\ref{fig:dists}. We find that energy dissipation is spread nearly uniformly across structures with $L$ and $W$ spanning a large range of scales. In particular, this range corresponds to inertial-range scales for $W$ and somewhat larger scales for $L$ (amplified by a factor of $B_0/b_\text{rms}$). In contrast, $E(T)T$ is peaked at $T$ deep within the dissipation range. Energy dissipation is peaked at smaller $T$ as~$Re$ is increased, consistent with a decreasing dissipation scale.

The dissipation-weighted distributions exhibit unambiguous scaling behavior with Reynolds number, with all distributions extending to smaller values with increasing $\operatorname{Re}$, consistent with a decreasing dissipation scale. We find that these dissipation-weighted distributions can be collapsed onto each other by considering the rescaled quantities $X_\text{Re} = X (\operatorname{Re}/\operatorname{Re}_0)^\gamma$ where $X \in \{L,W,T\}$, implying $X \sim \eta^{\gamma}$. Here, $\gamma$ is a scaling exponent (which may differ for the various scales) and $\operatorname{Re}_0=1000$ is an arbitrary reference Reynolds number. We find that the measurements are generally consistent with scaling indices in the range $1/2 < \gamma < 3/4$; in particular, the distributions for all three quantities can be rescaled reasonably well with $\gamma = 2/3$, as shown in the last row of Fig.~\ref{fig:dists}. The exponent $\gamma=2/3$ agrees with the perpendicular dissipation scale with scale-dependent dynamic alignment \citep{boldyrev_2006}, and is also inferred from the energy spectrum \citep{perez_etal_2014}. The results are also consistent with shallower scaling for the length than the width, which may be expected from critical balance, including $L \sim \eta^{1/2}$ and $W \sim \eta^{3/4}$ predicted for the parallel and perpendicular dissipation scales in the Goldreich-Sridhar phenomenology.

\begin{figure*}
\centering
\includegraphics[width=0.45\columnwidth]{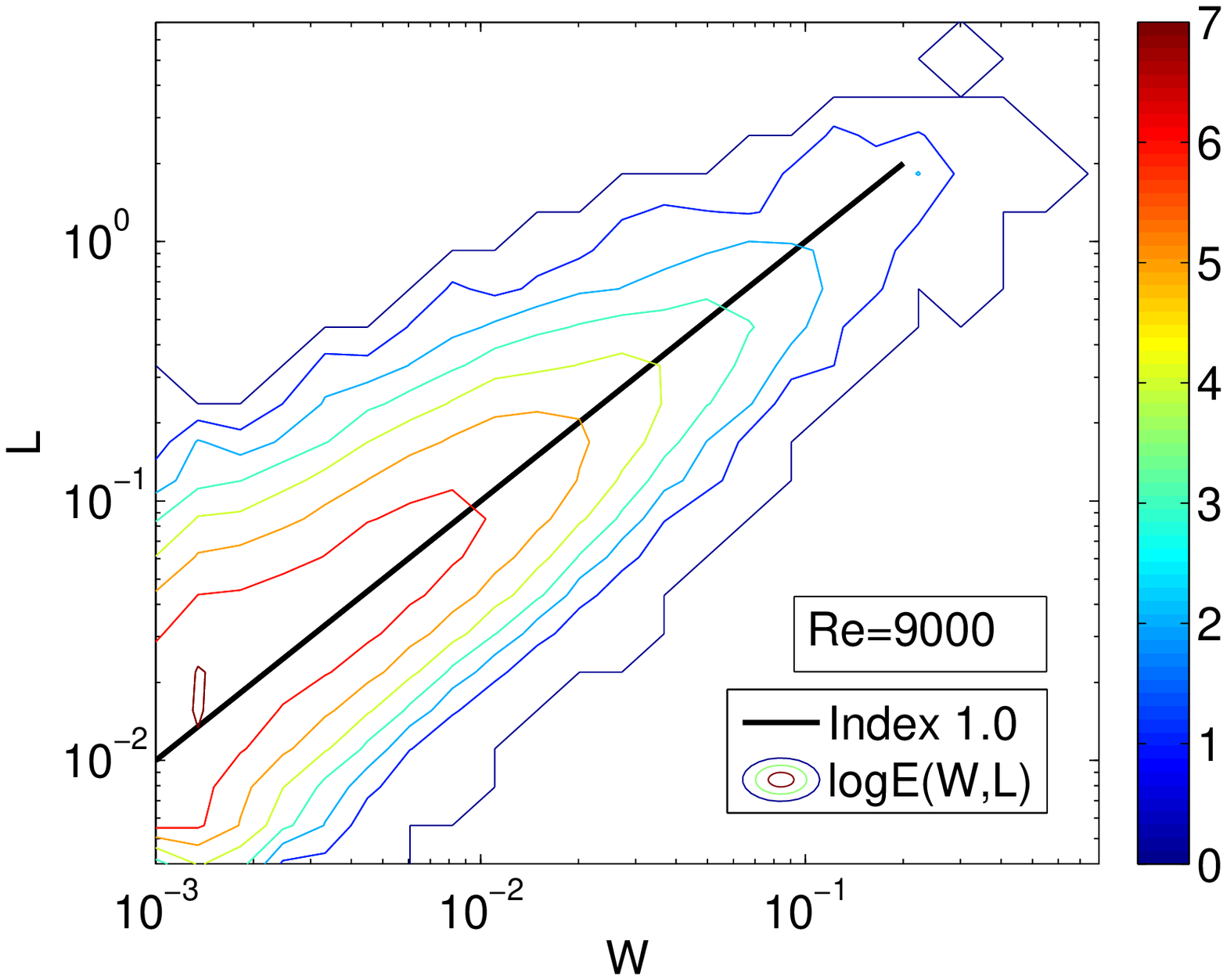}
\includegraphics[width=0.45\columnwidth]{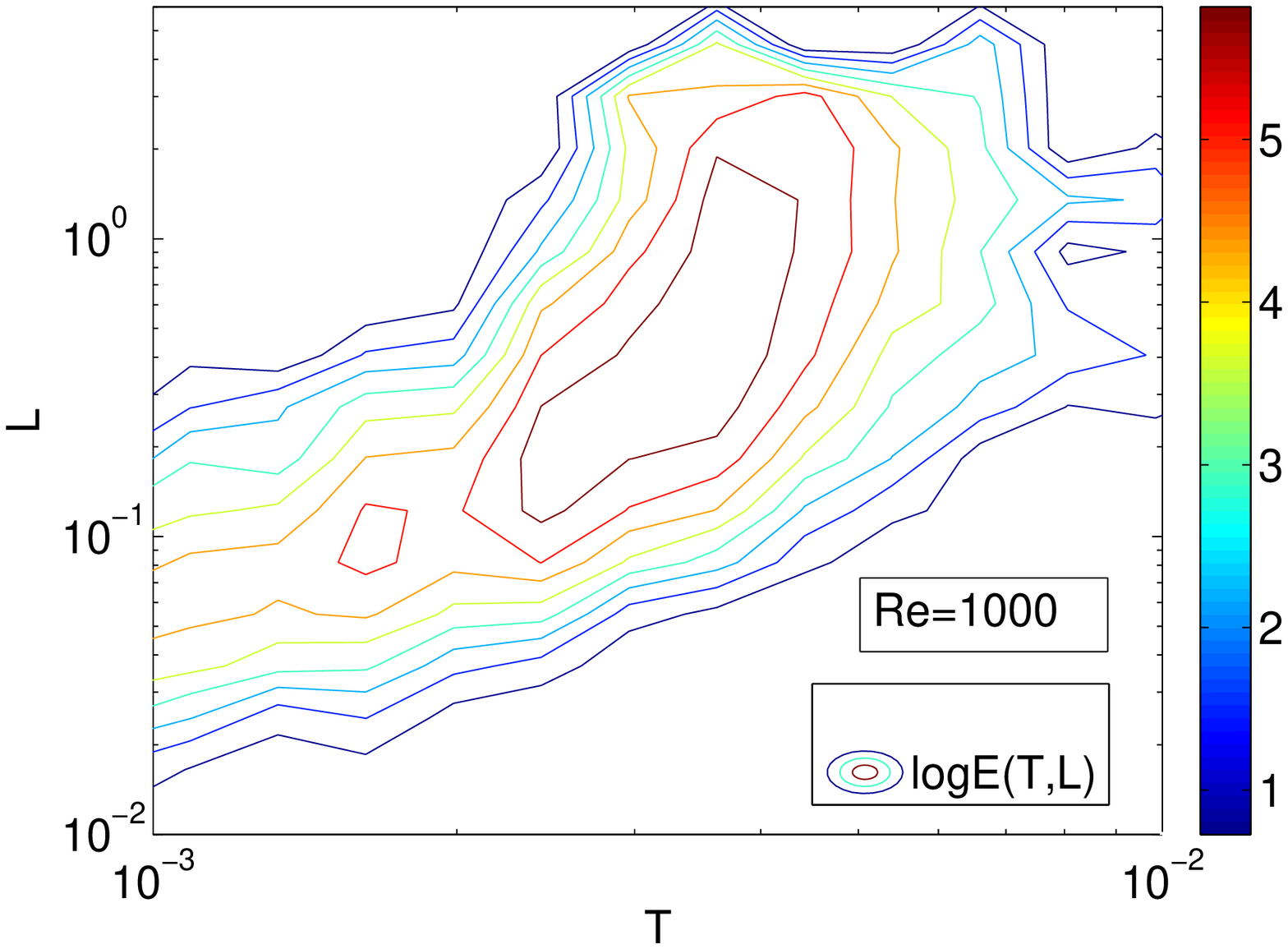}
\includegraphics[width=0.45\columnwidth]{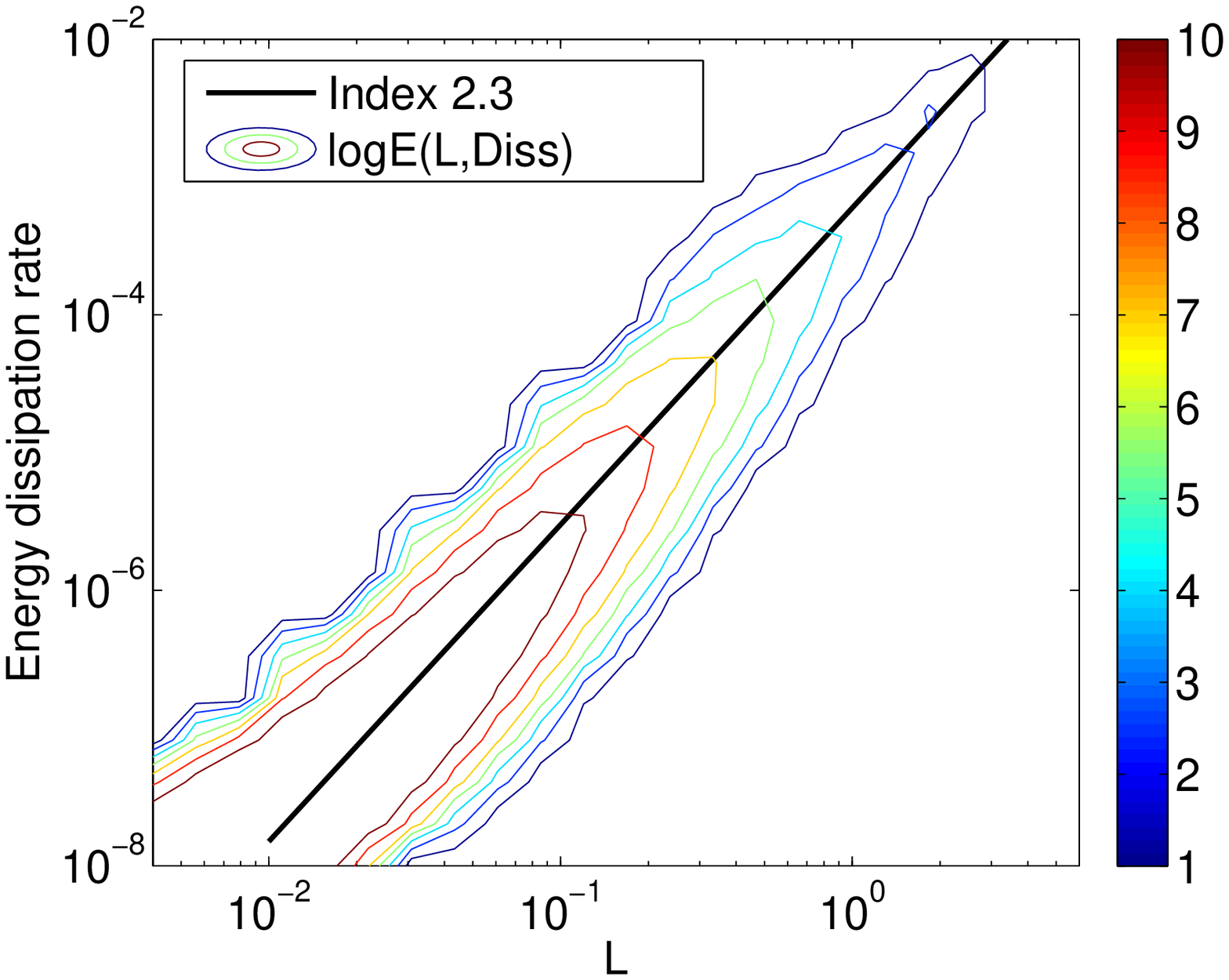}
 \caption{Contours of 2D dissipation-weighted probability distributions comparing length $L$ with width $W$ (top left) and thickness $T$ (top right). In order to obtain the largest scaling intervals, we show $\operatorname{Re} = 9000$ in the left plot and $\operatorname{Re} = 1000$ in the right plot. Also shown is the scaling of the dissipation rate ${\mathcal E}$ with $L$ (bottom). \label{fig:scalings2}}
 \end{figure*}

Finally, we consider the correlations between the different scales by plotting the 2D dissipation-weighted distributions $E(X,Y)$, where $E(X,Y) dX dY$ is the combined energy dissipation rate of all structures with scales $X$ and $Y$ (within bins of size $dX$ and $dY$). We show $E(W,L)$ and $E(T,L)$ in Fig.~\ref{fig:scalings2}, choosing $\operatorname{Re = 9000}$ and $\operatorname{Re = 1000}$, respectively, in order to obtain the largest scaling interval corresponding to the inertial range and dissipation range, respectively. There is a robust linear scaling between length and width, $L \approx 1.6 (B_0/b_\text{rms}) W$, in all of the simulations. In contrast, there is little to no correlation between $L$ and $T$ for large and intense structures, consistent with the thicknesses being fixed at the dissipation scale. Other scalings include the volume as $V \sim L^2$ (not shown) and energy dissipation rate as ${\cal E} \sim L^{2.3}$ (shown in Fig.~\ref{fig:scalings2}).

Although we focused our attention on the spatial characteristics of intermittent structures in this work, one can also consider the temporal characteristics by tracking the structures in time. We recently generalized the methodology applied in this work to perform a statistical analysis of 4D spatiotemporal structures in the current density\citep{zhdankin_etal_2015, zhdankin_etal_2015b}. We found that evolving structures have power-law distributions for the maximum quantities attained during their lifetimes (i.e., peak length, peak width, and peak energy dissipation rate) with similar indices as the purely spatial structures have for the corresponding instantaneous quantities. We also found the durations $\tau$ of evolving structures to be proportional to their maximum attained length, in agreement with critical balance ($\tau_\text{NL} \sim L/v_A$), which may be a robust phenomenon even inside of intermittent structures \citep[e.g.,][]{mallet_etal_2015}. We found the distribution of (time-integrated) dissipated energies to have a power law with index near $-1.75$, shallower than the critical index of $-2.0$ measured for the (instantaneous) energy dissipation rates in Fig.~\ref{fig:diss}. Incidentally, the distribution for dissipated energy is very similar to the observed distribution for energy released by solar flares \citep[][]{aschwanden_etal_2000b}. The indices for probability distributions and scaling relations inferred from the spatial and temporal analyses are compiled in Table~\ref{table-stats}, along with estimated error bars.

\begin{table*}[h!b!p!]
\caption{Scalings of current sheets in numerical simulations (* from temporal analysis \citep{zhdankin_etal_2015b}) \newline}
\begin{tabular}{|c|c|c|}
	\hline
\hspace{0.5 mm} Quantity \hspace{0.5 mm}  & \hspace{0.2 mm} Distribution index & \hspace{0.2 mm} Scaling index with $L$ \hspace{0.2 mm}  \\
	\hline
Dissipation rate ${\cal E}$ & $-2.0 \pm 0.1$ & $2.3 \pm 0.2$\\
Thickness $T$ & N/A & N/A \\
Width $W$ & $-3.3 \pm 0.1$ & $1.0 \pm 0.1$  \\
Length $L$ & $-3.3 \pm 0.1$ & $1.0$  \\
        \hline
Dissipated energy* & $-1.75 \pm 0.1$ & $3.0 \pm 0.2$  \\
Peak dissipation rate* & $-2.0 \pm 0.1$ & $2.0 \pm 0.2$ \\
Duration* & $-3.2 \pm 0.2$ & $1.0 \pm 0.1$ \\
	\hline
\end{tabular}
\centering
\label{table-stats}
\end{table*}

\section{Conclusions}

In this work, we presented an overview of the scaling properties of intermittent dissipative structures in driven incompressible MHD turbulence with a strong guide field. We found that the statistical properties of structures in the current density, vorticity, and Els\"{a}sser vorticity are nearly identical, despite the resistive and viscous contributions to the overall energy dissipation rate differing by a noticeable amount (with resistive dissipation exceeding viscous dissipation by about $50\%$). These sheet-like structures have lengths proportional to widths, $L \sim (B_0 / b_\text{rms}) W$, with both being distributed mainly across the inertial range. Thicknesses of the structures, on the other hand, are concentrated near the dissipation scale. When the temporal evolution of structures is accounted for, their durations are proportional to their maximum attained lengths, $\tau \sim L / v_A$. These scalings can be understood phenomenologically by focusing on Els\"{a}sser vorticity sheets: the RMHD equations imply that the lengths and widths may be associated with advection by the large-scale fields, while the thicknesses are associated with the balance between the nonlinear and dissipative terms. Since the two populations of Els\"{a}sser vorticity structures are only weakly correlated, it is no surprise that that structures in the current density and vorticity reflect the same statistical properties.

Intermittent current sheets may have observational consequences in a variety of astrophysical systems. They may explain high-energy flares in systems such as the solar corona \citep[e.g.,][]{dmitruk_gomez_1997, georgoulis_2005, uritsky_etal_2007}, pulsar wind nebulae, accretion disks, and jets. Indeed, as described in our temporal analysis\citep{zhdankin_etal_2015b}, the statistical properties of dissipative events in MHD turbulence are largely consistent with solar flare observations. Intermittent current sheets also naturally explain magnetic discontinuities and energetic particles in the solar wind \citep{bruno_etal_2001, greco_etal_2010} and Earth's magnetosphere \citep{angelopoulos_etal_1999}, although it may be nontrivial to infer the sizes of the structures from measurements taken by a single spacecraft. 

There remain a number of important open questions. Firstly, how does changing the magnetic Prandtl number or including more realistic mechanisms of dissipation affect intermittency? In particular, how does this change the relative roles of magnetic and kinetic dissipation\citep{brandenburg_2014, sahoo_etal_2011}, characteristic thicknesses, distribution of energy dissipation rates, and other statistical properties of structures in the various intermittent fields? We note that since many astrophysical plasmas are collisionless, a kinetic framework is required to self-consistently describe the small-scale dynamics, including mechanisms of energy dissipation and particle acceleration. Progress may be spurred by recent simulations of collisionless plasma turbulence that show signatures of intermittency at scales near and below the ion gyroradius \citep{wan_etal_2012b, leonardis_etal_2013, karimabadi_etal_2013, tenbarge_howes_2013}, although larger simulations may be needed to properly describe the inertial-range deposition of energy to small scales \citep{parashar_etal_2015}. Measurements of structure functions provide provocative hints that kinetic scales in the solar wind may be scale invariant\citep{kiyani_etal_2009}, but it is unclear whether this is a generic outcome and what implications it has for the existence of coherent structures.

Secondly, how are intermittent structures affected by large-scale conditions such as background flows, boundary conditions, and driving mechanisms? We anticipate that, in some cases, the large-scale conditions may limit the size of the largest structures. Turbulence in line-tied MHD \citep{ng_bhattacharjee_1998, wan_etal_2014, pongkitiwanichakul_etal_2015} or rotating shear flows, where the magnetorotational instability operates\citep{balbus_hawley_1998}, could be more realistic settings for astrophysical systems such as stellar coronae and accretion disks, respectively. Conversely, it is conceivable that the lengths and durations of structures may grow to exceed the driving scale in unbounded geometries \citep{dmitruk_matthaeus_2007}.

Thirdly, do the large intermittent structures discussed in this work remain stable for arbitrarily high fluid and magnetic Reynolds numbers? There are hints that vorticity filaments in hydrodynamic turbulence become unstable at high Reynolds numbers, being supplanted by cloud-like clusters of structures \citep{ishihara_etal_2009}. One may likewise expect on general grounds that current sheets become unstable after reaching a critical aspect ratio due to tearing or Kelvin-Helmholtz instabilities \citep{loureiro_etal_2007, loureiro_etal_2013}. However, it is unclear how to model the background turbulent flow to make concrete predictions on the stability of intermittent structures. Instabilities may set an upper limit on the sizes and dissipation rates of structures, and may impart signatures on their temporal evolution. If intermittent structures lose their coherence at large $\operatorname{Re}$, then a more general methodology to characterize the clustering of small structures may be pursued. On a related note, measuring the spatial correlations between current sheets, vorticity sheets, and other quantities may give insights for modeling the structures.

Fourthly, how is intermittency manifest in other quantities such as magnetic fields, density, temperature, and nonthermal particle acceleration? Intermittent magnetic fields may develop as a consequence of the magnetic dynamo, possibly explaining filamentary structures in the solar photosphere \citep{cattaneo_etal_2003} and the center of the Galaxy \citep{boldyrev_yusef-zadeh_2006}. Intermittent density or temperature profiles may affect radiative characteristics and chemical processes, being relevant for compressible turbulence in the interstellar medium \citep{boldyrev_etal_2002b, falgarone_etal_2015}, deflagration in supernovae \citep{pan_etal_2008}, and heating in the solar corona \citep{dahlburg_etal_2012}. It is unclear to what extent intermittency may affect these quantities in different systems.

In conclusion, there is much to be discovered about turbulence and its intermittency in a wide variety of settings, including the relatively simple case of reduced MHD considered here. A complete understanding requires one to explore beyond the energy spectrum and low-order statistics, and instead scrutinize the higher-order statistics and morphology of the turbulent fields. We hope that the insights from the analysis presented here will provide guidance for future studies of intermittency in MHD turbulence and beyond.

\acknowledgements

The authors would like to thank Jean Carlos Perez for support with the numerical results described in this paper and for performing the larger simulations. This research was supported by the NSF Center for Magnetic Self-Organization in Laboratory and Astrophysical Plasmas at the University of Wisconsin-Madison. SB was supported by the Space Science Institute, NSF grant AGS-1261659, and NASA grant NNX14AH16G. DU was supported by NASA grant No. NNX11AE12G, US DOE grants DE-SC0008409, DE-SC0008655, and NSF grant AST-1411879.

\bibliography{refs_pop}

\end{document}